\documentclass[a4paper,fleqn,usenatbib]{mnras}

\usepackage{newtxtext,newtxmath}

\usepackage[T1]{fontenc}
\usepackage{ae,aecompl}

\usepackage{graphicx}	
\usepackage{amsmath}	
\usepackage{amssymb}	
\usepackage{array}

\newcolumntype{M}[1]{>{\centering\arraybackslash}m{#1}}

\title[Neutron star jets]{Disk--jet coupling in low--luminosity accreting neutron stars}

\author[V. Tudor]{V. Tudor,$^1$\thanks{E-mail: vlad.tudor@icrar.org}
J. C. A. Miller-Jones,$^1$
A. Patruno,$^{2,3}$
C. R. D'Angelo,$^{2}$
\newauthor
P. G. Jonker,$^{4,5}$
D. M. Russell,$^{6}$
T. D. Russell,$^{1,7}$
F. Bernardini,$^{6}$
F. Lewis,$^{8,9}$
\newauthor
A. T. Deller,$^{10}$
J. W. T. Hessels,$^{3,7}$
S. Migliari,$^{11,12}$
R. M. Plotkin,$^{1}$
R. Soria$^{1}$ and
\newauthor
R. Wijnands$^{7}$
\\
\\
$^1$International Centre for Radio Astronomy Research -- Curtin University, GPO Box U1987, Perth, WA 6845, Australia\\
$^2$Leiden Observatory, Leiden University, PO Box 9513, NL-2300 RA
Leiden, the Netherlands\\
$^3$ASTRON, the Netherlands Institute for Radio Astronomy, Postbus 2, 7990 AA, Dwingeloo, the Netherlands\\
$^4$SRON, Netherlands Institute for Space Research, Sorbonnelaan 2, 3584~CA, Utrecht, the Netherlands.\\
$^5$Department of Astrophysics/IMAPP, Radboud University, P.O.~Box 9010, 6500 GL, Nijmegen, the Netherlands.\\
$^6$New York University Abu Dhabi, PO Box 129188, Abu Dhabi, UAE\\
$^7$Anton Pannekoek Institute for Astronomy, University of Amsterdam, PO Box 94249, NL-1090 GE Amsterdam, the Netherlands\\
$^8$Faulkes Telescope Project, School of Physics, and Astronomy, Cardiff University, The Parade, Cardiff CF24 3AA, UK\\
$^9$Astrophysics Research Institute, Liverpool John Moores University, 146 Brownlow Hill, Liverpool L3 5RF, UK\\
$^{10}$Centre for Astrophysics and Supercomputing, Swinburne University of Technology, PO Box 218, Hawthorn, VIC 3122, Australia\\
$^{11}$European Space Astronomy Centre (ESAC), E-28692 Villanueva de la Ca{\~n}ada, Madrid, Spain\\
$^{12}$Department of Quantum Physics and Astrophysics \& Institute of Cosmos Sciences, University of Barcelona, E-08028 Barcelona, Spain
}

\date{Accepted XXX. Received YYY; in original form ZZZ}

\pubyear{2017}

\begin{document}
\label{firstpage}
\pagerange{\pageref{firstpage}--\pageref{lastpage}}
\maketitle


\begin{abstract}
In outburst, neutron star X-ray binaries produce less powerful jets than black holes at a given X-ray luminosity. This has made them more difficult to study as they fade towards quiescence. To explore whether neutron stars power jets at low accretion rates ($L_{\rm X} \lesssim 10^{36}$\,erg\,s$^{-1}$), we investigate the radio and X-ray properties of three accreting millisecond X-ray pulsars (IGR\,J17511--3057, SAX\,J1808.4--3658 and IGR J00291+5934) during their outbursts in 2015, and of the non-pulsing neutron star Cen\,X--4 in quiescence (2015) and in outburst (1979). We did not detect the radio counterpart of IGR\,J17511--3057 in outburst or of Cen\,X--4 in quiescence, but did detect IGR\,J00291+5934 and SAX\,J1808.4--3658, showing that at least some neutron stars launch jets at low accretion rates. While the radio and X-ray emission in IGR\,J00291+5934 seem to be tightly correlated, the relationship in SAX\,J1808.4--3658 is more complicated. We find that SAX\,J1808.4--3658 produces jets during the reflaring tail, and we explore a toy model to ascertain whether the radio emission could be attributed to the onset of a strong propeller. The lack of a universal radio/X-ray correlation, with different behaviours in different neutron star systems (with various radio/X-ray correlations; some being radio faint and others not), points at distinct disk--jet interactions in individual sources, while always being fainter in the radio band than black holes at the same X-ray luminosity.
\end{abstract}




\section{Introduction}

Radio jets are a common feature among accreting compact objects, most prominently associated with black holes \citep[e.g.][]{1994Natur.371...46M}, but also produced by neutron stars \citep{2001ApJ...553L..27F} and white dwarfs \citep{2015MNRAS.451.3801C}. Similar to black holes, neutron star X-ray binaries have been found to produce both resolved and compact radio jets \citep{2004Natur.427..222F, 2010ApJ...716L.109M, 2010ApJ...710..117M, 2013MNRAS.435L..48S}, albeit at a fainter radio luminosity than black holes at the same X-ray luminosity in the hard state \citep[a factor of $\approx$ 30;][]{2006MNRAS.366...79M}, when the X-ray spectrum is dominated by a hard power-law component \citep{1998ASIC..515..279V}. 

Despite a long history of observations, it is still unknown how jets are launched. Comparing different classes of accreting objects will provide clues regarding which jet properties are universal, and which rely on the nature of the compact object. For example, comparing neutron star and black hole jets can help distinguish between different jet models and jet launching mechanisms. Possibilities include the Blandford--Payne model (whereby the jet is produced by the collimation of disk winds by the magnetic fields in the disk, and is possible in both neutron stars and black holes; \citealp{1982MNRAS.199..883B}), the Blandford--Znajek model (in which the jet is launched by the magnetic field of the disk, extracting angular momentum from the ergosphere of a spinning black hole; \citealp{1977MNRAS.179..433B}), or models powered by the spin of a neutron star, such as the X-wind model (initially developed for young stellar objects, \citealp{1994ApJ...429..781S}; but potentially applicable to neutron stars; \citealp{2006MNRAS.366...79M, 2012IJMPS...8..108M}).

One way to compare neutron star and black hole jets is through investigations of their disk--jet  coupling. The connection between inflow (disk) and outflow (jets) processes is manifested in the empirical radio\,--\,X-ray luminosity correlation of hard-state black hole X-ray binaries, which links the radio luminosity (set by the jet power) with the X-ray luminosity \citep[a proxy for the mass accretion rate through the inner regions of the accretion flow;][]{2003MNRAS.343L..99F}. This relationship takes the form of a power-law between the radio and X-ray luminosities, $L_{\rm R} \propto L_{\rm X}^\beta$, where $\beta \approx 0.5 - 0.7$ in the most well-studied black hole sources such as GX\,339--4 or V404\,Cygni \citep{2000A&A...359..251C, 2003MNRAS.344...60G, 2013MNRAS.428.2500C, 2014MNRAS.445..290G, 2017ApJ...834..104P}.

Knowledge of the difference between the radio emission of neutron stars and black holes can also be used to determine the nature of the accreting compact object in new transient or quiescent systems. The three ways of irrefutably distinguishing between neutron stars and black holes are the detection of Type~I X-ray bursts \citep{1977Natur.270..310J}, coherent X-ray pulsations \citep{1973ApJ...184..271L}, which both imply that the compact object is a neutron star, or the measurement of the mass function of the system in question \citep{2007IAUS..238....3C}. Due to the low brightness of donor stars, and the large distances and high extinctions typical of X-ray binaries, mass functions can often be very difficult to measure. So, in the absence of Type~I bursts or pulsations, it is difficult to identify of the nature of the accretor. One promising approach is through simultaneous radio and X-ray observations, especially since radio facilities have reached the requisite sensitivity to detect an increasing number of X-ray binaries \citep[e.g. the Karl G. Jansky Very Large Array;][]{2011ApJ...739L...1P}, and to track the closest systems as they decay back towards quiescence.

Transient X-ray binaries undergo outbursts in which their X-ray luminosities increase by up to six orders of magnitude over their quiescent levels, accompanied by spectral changes and strong outflows. As radio jets from black hole X-ray binaries in outburst are bright, a factor of at least 30 louder than those of ``ordinary'' neutron stars \citep{2006MNRAS.366...79M}, it has been possible to study the radio--X-ray correlation for black holes over 8 orders of magnitude in X-ray luminosity. As systems change luminosities in the hard state as they decay towards quiescence on week to month timescales, inflow--outflow coupling can be directly traced in individual sources \citep{2008MNRAS.389.1697C, 2013MNRAS.428.2500C, 2011IAUS..275..255C, 2017ApJ...834..104P}. 

\citet{2014MNRAS.445..290G} found that the radio--X-ray correlations of individual black hole systems have less scatter than the black hole X-ray binary population as a whole. Different correlation slopes $\beta$ in different sources suggest different, or evolving, radiative efficiencies \citep{2003MNRAS.343L..59H}. Above $L_{\rm X} \approx 10^{36}$\,erg\,s$^{-1}$, a few black hole X-ray binaries are seen to follow a steeper ($\beta \approx 1.4$), radiatively efficient track in the radio/X-ray plane \citep{2011IAUS..275..255C}. As they become fainter, the radiatively efficient systems that have been observed long enough, are seen to switch to the primary, radiatively inefficient track with $\beta \approx 0.7$ (H\,1743--322, \citealp{2010MNRAS.401.1255J}; XTE\,J1752--223, \citealp{2012MNRAS.423.2656R}; MAXI\,J1659--152, \citealp{2012MNRAS.423.3308J}). \citet{2014MNRAS.445..290G}, however, did not find strong statistical evidence for the existence of two tracks in the overall black hole population, meaning that individual sources could follow different tracks.

In comparison to black holes, the inflow--outflow connection in neutron stars is complicated by the presence of a crust, a magnetosphere and possibly other as-yet unknown factors. Specifically, the crust prevents accreted material from reaching an ergosphere, the magnetic field can alter, or halt the accretion flow and outflows through mechanisms that do not operate for black holes \citep[such as the propeller effect;][]{1975A&A....39..185I}. In addition, due to their less luminous jets, a narrower range of radio luminosities are accessible for neutron stars before telescope sensitivity limits are reached. Because of this, neutron star jets have been less well explored to date, with few reported radio detections for systems below $L_{\rm X} \approx 10^{36}$\,erg\,s$^{-1}$.

In the case of black holes, the radio\,--\,X-ray correlation is only seen in the hard and quiescent X-ray states. In the soft state, at higher accretion rates, when the spectrum is dominated by a soft, thermal component, steady jets are suppressed in black holes \citep{2004MNRAS.355.1105F} and some neutron stars \citep{2003MNRAS.342L..67M, 2009MNRAS.400.2111T, 2010ApJ...716L.109M, 2011IAUS..275..233M}, although \citet{2004MNRAS.351..186M} present two radio detections of a neutron star in the soft state. The analogous low-luminosity, hard-state neutron star systems are the atoll sources (ordinary neutron stars that are typically accreting at 1--10\% the Eddington luminosity ($L_{\rm Edd}$); \citealp{2002ApJ...568L..35M}), accreting millisecond pulsars (AMXPs), and transitional millisecond pulsars (tMSPs).  In the context of the ``recycling'' scenario \citep{1974SvA....18..217B, 1982Natur.300..728A, 1982CSci...51.1096R}, AMXPs and tMSPs are thought to be missing links between radio millisecond pulsars and neutron star X-ray binaries. AMXPs display X-ray pulsations at the spin period of the neutron star \citep{1998Natur.394..344W}, whereas tMSPs switch between an accreting state, when the system is observed as an X-ray binary, and a non-accreting state, when the neutron star becomes a radio millisecond pulsar \citep{2009Sci...324.1411A}. The tMSP M28I behaved as an AMXP in outburst \citep{2013Natur.501..517P}, and the tMSPs PSR\,J1023+0038 \citep{2015ApJ...807...62A} and XSS\,J12270--4859 \citep{2015MNRAS.449L..26P} exhibit coherent X-ray pulsations akin to AMXPs, so tMSPs may in fact be a subset of AMXPs. So far, studies hint at diverse disk--jet coupling relations for different neutron stars, as described below.

The first studies that showed disk--jet coupling in neutron stars focused on sources which persistently accrete close to the Eddington luminosity. Their radio emission appeared to correlate with the spectral state of the source \citep{1988Natur.336..146P, 1990A&A...235..147H, 1990ApJ...365..681H}. \citet{2003MNRAS.342L..67M} later found other evidence for disk--jet coupling in the source 4U\,1728--34, which follows a relationship of the form $L_{\rm R} \propto L_{\rm X}^{1.5 \pm 0.2}$, as theoretically predicted for radiative efficient accretion flows. \citet{2006MNRAS.366...79M} extended their sample with three additional atolls in the hard state, which, as a sample, displayed a similar $\beta \approx 1.4$ relationship. In addition, they also added four AMXPs to their sample, showing that AMXPs are scattered more on the radio\,--\,X-ray luminosity plane, and can reach slightly higher radio luminosities than atoll sources at similar X-ray luminosities. Later, the atoll source Aql\,X--1 was found to follow its own correlation \citep{2009MNRAS.400.2111T}, with a coupling index of $\beta = 0.76 \pm 0.15$ when only the hard states are considered \citep{2016MNRAS.tmp..785T}. The neutron star transient EXO\,1745--248 in the globular cluster Terzan~5 traced a $\beta=1.7 \pm 0.1$ correlation during its 2015 outburst, albeit a factor of five less luminous in the radio band than 4U\,1728--34 and Aql\,X--1 \citep{2016MNRAS.tmp..785T}, suggesting different normalizations for different sources. The measurement of correlation normalizations, however, can be biased by incorrect distance estimates \citep{2004MNRAS.351.1359J}. They also suggested that tMSPs may trace a radio\,--\,X-ray path of the same slope as black holes, but with a lower normalisation, having radio jets that are five times fainter than black holes but one order of magnitude brighter than ordinary neutron stars. \citet{2015ApJ...809...13D} suggested the relative radio loudness of tMSPs could be produced by a radiatively-inefficient flow, possibly due to the ejection of matter in the propeller regime.

With the exception of the tMSPs, which have radio detections below $L_{\rm X} \lesssim 10^{36}$\,erg\,s$^{-1}$, the majority of radio detections in the above studies are of neutron stars in the range $L_{\rm X} = 10^{36}-10^{37}$\,erg\,s$^{-1}$. This makes the measure of the correlation index $\beta$ uncertain due to the short lever arm in X-ray luminosity. X-ray luminosities should be sampled over more than two orders of magnitude to ensure that the measured $\beta$ does not sample temporary deviations from the underlying correlation \citep{2013MNRAS.428.2500C}. Here, we report new and archival radio and X-ray observations of four neutron star X-ray binaries, carried out to track jet behaviour below $L_{\rm X} = \rm 10^{36}\,erg\,s^{-1}$. Our target sample consists of \mbox{Cen\,X--4}, it being the closest known quiescent neutron star low mass X-ray binary (1.2\,kpc), and the AMXPs IGR\,J17511--3057, SAX\,J1808.4--3658 and IGR\,J00291+5934, which all underwent outbursts in 2015. With this sample, we can track the radio--X-ray behaviour of ordinary neutron stars and AMXPs and compare them to tMSP and black hole systems.

\section{Targets}
\label{sec:targets}

\subsection{Quiescent neutron star transient}
\subsubsection{Cen\,X--4}

\mbox{Cen\,X--4}, a likely atoll source, was discovered during an outburst in 1969 \citep{1969ApJ...157L.157C, 1970ApJ...159L..57E}. During a subsequent outburst in 1979 \citep{1980ApJ...241..779K}, it was also detected in the radio band by the VLA, which was being built at the time \citep{1979IAUC.3369....1H, 1988ApJ...335L..75H}. \mbox{Cen\,X--4} has been in quiescence ever since, although the currently variable X-ray emission indicates that it is still actively accreting, albeit at a very low level \citep[$L_{\rm X} = 10^{32}$\,erg\,s$^{-1}$;][]{2000A&A...358..583C, 2004ApJ...601..474C, 2010ApJ...720.1325C, 2013MNRAS.436.2465B}. Given the radio detection of the tMSP PSR\,J1023+0038 in its accretion state \citep{2015ApJ...809...13D}, during which it has a similar X-ray luminosity as Cen\,X--4 in quiescence, and indications of a jet in the quiescent AMXP XTE\,J1814--338 \citep{2013A&A...559A..42B}, we hypothesise that a jet could still be launched in \mbox{Cen\,X--4} even at extremely low X-ray luminosities ($L_{\rm X} \approx 10^{-6} \: L_{\rm Edd}$).

Based on the theoretical Eddington luminosity for a neutron star, the Type I X-ray bursts of \mbox{Cen\,X--4} place it at a distance of $1.2 \pm 0.3$\,kpc \citep{chevalier89}. Using the empirical Eddington luminosity calibrated for neutron stars in globular clusters, \citet{2005A&A...435.1185G} report a similar distance of $1.4 \pm 0.3$\,kpc. For consistency with other studies of \mbox{Cen\,X--4}, we adopt a distance of 1.2\,kpc.

As the most nearby known neutron star X-ray binary, \mbox{Cen\,X--4} is a good laboratory for testing whether ordinary neutron stars launch jets in quiescence. Previous radio observations of the quiescent \mbox{Cen\,X--4}, however, have failed to detect a jet, with a 4$\sigma$ upper limit of $\approx 0.4$\,mJy \citep{1992xbfb.work...99K}. This upper limit ($L_{\rm R} < 3 \times 10^{27}$\,erg\,s$^{-1}$) is unconstraining compared to black holes at the same X-ray luminosity ($L_{\rm R} \approx 10^{27}$\,erg\,s$^{-1}$).

\subsection{AMXPs in outburst}
\subsubsection{IGR\,J17511--3057}
\label{sec:intr_j17511}

Discovered in 2009 by \textit{INTEGRAL} \citep{2009ATel.2196....1B}, IGR\,J17511--3057 was later classified as an AMXP after the detection of 245\,Hz X-ray pulsations \citep{2009ATel.2197....1M}. Since its 2009 outburst, it has undergone two additional outbursts, in 2012 and 2015. Previous radio observations were conducted by \citet{2009ATel.2232....1M} during the September 2009 outburst, but no emission was detected, with 3$\sigma$ upper limits of 0.17\,mJy.

Based on the peak X-ray flux of its thermonuclear bursts, \citet{2010MNRAS.409.1136A} placed an upper limit of 6.9\,kpc on its distance. \citet{2010MNRAS.407.2575P} found a similar lower limit of 6.5\,kpc by assuming that the emission during the outburst decay comes from the neutron star surface. The pulsation timing analysis of \citet {2011A&A...526A..95R} pointed towards two possible distances (6.3 or 3.6\,kpc), the first of which is comparable with previous estimates. We adopt a distance of 6.9\,kpc for IGR\,J17511--3057 for consistency with previous studies.

\subsubsection{SAX\,J1808.4--3658}

The first detected and most extensively studied AMXP, SAX\,J1808.4--3658 was discovered in 1996 with the \textit{BeppoSAX} satellite \citep{1998A&A...331L..25I}, and classified as an AMXP after the discovery of a 401\,Hz spin \citep{1998Natur.394..344W}. It was previously detected in the radio band during three of its outbursts, in 1998, 2002 and 2005 \citep{1999ApJ...522L.117G, 2002IAUC.7997....2R, 2005ATel..524....1R}.

Based on the fluence and recurrence times of its outbursts, \citet{2006ApJ...652..559G} estimated a distance to SAX\,J1808.4--3658 of 3.4--3.6\,kpc, which is consistent with the distance derived from Type I X-ray bursts (3.1--3.8\,kpc). We adopt a distance of 3.5\,kpc for SAX\,J1808.4--3658.

The X-ray light curves of SAX\,J1808.4--3658 are similar between different outbursts. After a quick rise to the outburst peak, the source slowly decays for 15--20 days, followed by a faster decay over $\approx$ 3 days, after which the source reflares for a few tens of days, each reflare lasting a few days \citep{2016ApJ...817..100P}. Episodic accretion of matter in the propeller regime could be the mechanism behind the reflares \citep{2016ApJ...817..100P}. At $L_{\rm X} \approx 10^{33}$\,erg\,s$^{-1}$, the magnetosphere reaches the light cylinder of the neutron star, so that in quiescence SAX\,J1808.4--3658 should turn on as a radio pulsar \citep{2002ApJ...575L..15C}. However, no radio pulsations have been detected so far \citep{2003ApJ...589..902B, 2010A&A...519A..13I, 2016arXiv161106023P} in quiescence \citep[quiescent $L_{\rm X} \approx 10^{31}$\,erg\,s$^{-1}$;][]{2002ApJ...575L..15C}, when a typical radio pulsar should have been detected, although the presence of a radio pulsar has been indirectly inferred \citep{2003A&A...404L..43B, 2004ApJ...614L..49C}.

\subsubsection{IGR\,J00291+5934}

Following the discovery of IGR\,J00291+5934 in 2004 by \textit{INTEGRAL} \citep{2004ATel..352....1E}, the detection of a 599\,Hz spin frequency by \textit{RXTE} classified it as an AMXP \citep{2005ApJ...622L..45G}. Similar to SAX\,J1808.4--3658, IGR\,J00291+5934 goes into outburst every 3--4 years. During its discovery outburst, the source was also detected in the radio band \citep{2004ATel..355....1P, 2004ATel..361....1F, 2004ATel..364....1R}, but no radio emission was detected during its two fainter outbursts in 2008 \citep{2008ATel.1667....1L, 2010A&A...517A..72L}. Similar to SAX\,J1808.4--3658, there is some evidence that IGR\,J00291+5934 might become active as a radio pulsar during quiescence \citep{2008ApJ...680..615J}.

A Type I X-ray burst from IGR\,J00291+5934 was detected for the first time during the 2015 outburst, placing it at a distance of $4.2\pm0.3$\,kpc \citep{2015ATel.7852....1B, 2017A&A...599A..88D}. This measurement is in line with the lower limit of $\approx$ 4\,kpc based on the fluence of the 2004 outburst \citep{2005ApJ...622L..45G}. However, \citet{2005MNRAS.361..511J} and \citet{2008ApJ...672.1079T} found that IGR\,J00291+5934 should lie 2--3.6\,kpc away, assuming a similar quiescent X-ray flux as other AMXPs. \citet{2008ApJ...672.1079T} also estimated a lower limit of 1.8--3.8\,kpc based on the timing of the fast decay in its X-ray light curve, and assumptions on disk ionization states. In this work, we assume a distance of 4.2\,kpc to IGR\,J00291+5934, since it is based on Type I X-ray bursts.

\section{Observations}

\subsection{Cen\,X--4}

On 2015 January 17, we observed \mbox{Cen\,X--4} in quiescence simultaneously with the Karl G. Janksy Very Large Array (VLA, program 14B--117) and the \textit{Swift} X-ray Telescope \citep[XRT;][]{2004ApJ...611.1005G, 2005SSRv..120..165B}, under a Target of Opportunity observation (ObsID 00035324066). The VLA observations were carried out when the array was in the CnB configuration, at 8--12\,GHz, between 13:06 -- 13:48 UT, with an on-source integration time of 34 minutes. The XRT observations started at 13:25 UT, for a deadtime-corrected exposure time of 936\,s in photon counting mode.

\subsection{IGR\,J17511--3057}

An outburst from IGR\,J17511--3057 was detected by \textit{INTEGRAL} on 2015 March 23 \citep{2015ATel.7275....1B, 2016A&A...596A..71P}. We observed this source with the VLA (program 14B--153) on three occasions over two weeks, during a period of X-ray monitoring with \textit{Swift}/XRT (detailed in \citealp{2016A&A...596A..71P}). Our radio observations were carried out when the VLA was in B configuration, at 8--12\,GHz. Because IGR\,J17511--3057 remained undetected in all these observations, we then switched to observing SAX\,J1808.4--3658, which went into outburst during our monitoring campaign on IGR\,J17511--3057.

\subsection{SAX\,J1808.4--3658}

The \textit{Swift}/BAT monitor detected a new outburst from SAX\,J1808.4--3658 on 2015 April 9 \citep{2015ATel.7364....1S}, which we subsequently monitored with the VLA (program 14B--153) in B configuration, during the decay phase in April and May. On April 19, we observed this source simultaneously in the C (4--8\,GHz) and K (18--26.5\,GHz) bands, for improved spectral information at peak luminosity. The high-frequency K band observations required pointing calibration, for which we observed the bright calibrator J1820--2528. All the other observations were taken in the X (8--12\,GHz) band. The outburst was monitored in X-rays with \textit{Swift}/XRT with almost daily cadence (the observation log is given in Table~\ref{tab:xlog1808}). For comparison with the X-ray and radio data, we also report optical observations; Bernardini et al., in preparation).

We monitored the 2015 outburst of SAX J1808.4--3658 at optical wavelengths with the 2-m robotic Faulkes Telescopes North and South, located at Haleakala on Maui and Siding Spring, Australia, respectively, as part of an ongoing monitoring campaign of $\sim 40$ low-mass X-ray binaries \citep{lewis08}. During the outburst we increased the cadence and made additional observations with five of the robotic 1-m telescopes in the Las Cumbres Observatory (LCO) network; namely one located at the Cerro Tololo Inter-American Observatory, Chile, two at the South African Astronomical Observatory, Sutherland, South Africa, and two at Siding Spring Observatory, Australia. Photometric observations were made using four filters; Bessell $B$-band, $V$-band, $R$-band and Sloan Digital Sky Survey (SDSS) $i^{\prime}$-band. Here, we present the data in $i^{\prime}$-band only, which have the most number of observations during the outburst, and are sufficient to characterise the optical light curve, identify flares, and compare to the X-ray and radio light curves (Fig.~\ref{fig:j1808_lc}).

\begin{table*}
  \caption{Swift/XRT observations of SAX\,J1808.4--3658 in window timing (WT) and photon counting (PC) modes.}
  \centering
  \begin{tabular*}{0.7\textwidth}{@{\extracolsep{\fill}}lllll}
    \hline
    \hline
Program IDs & Obs. Number & Obs. Mode & Date & Exposure (s)\\
\hline
30034 & 74 & WT & 2015-04-13 & 5184 \\
      & 76 & WT & 2015-04-14 & 1618 \\
      & 77 & WT & 2015-04-15 & 839  \\
81453 & 01 & WT & 2015-04-16 & 1794 \\
33737 & 01 & WT & 2015-04-17 & 5179 \\
30034 & 78 & WT & 2015-04-18 & 1848\\
& 79 & WT & 2015-04-19 & 389 \\
& 80 & WT & 2015-04-20 & 621 \\
& 81 & WT & 2015-04-21 & 1969\\
& 82 & WT & 2015-04-22 & 2188\\
33737 & 02 & PC & 2015-04-24 & 3719\\
& 03 & PC & 2015-04-29 & 2544\\
& 04 & PC & 2015-04-30 & 5444\\
33034 & 83 & PC & 2015-05-02 & 418\\
& 84 & PC & 2015-05-03 & 1118\\
& 85 & PC & 2015-05-04 & 1003\\
& 86 & PC & 2015-05-05 & 684\\
& 87 & PC & 2015-05-06 & 1118\\
33737& 05 & PC & 2015-05-09 & 5855\\
30034 & 88 &PC & 2015-05-09 & 975\\
& 90 & PC & 2015-05-10 & 1041\\
& 89 & PC & 2015-05-10 & 1076\\
& 91 & PC & 2015-05-12 & 1166\\
& 92 & PC & 2015-05-13 & 968\\
33737& 06 & PC & 2015-05-15 & 3707\\
30034& 94 & PC & 2015-05-19 & 995\\
\hline
  \end{tabular*}
  \label{tab:xlog1808}
\end{table*}

\subsection{IGR\,J00291+5934}

Following the optical discovery of a new outburst from IGR\,J00291+5934 on 2015 July 24 by \textit{Swift}/BAT \citep{2015GCN..18051...1C}, in August we carried out three radio observations with the VLA in A configuration, at 8--12\,GHz, under Director's Discretionary Time (program 15B--339).

\begin{table*}
	\centering
	\begin{minipage}[b]{0.88\textwidth}
	\caption{Summary of VLA radio observations.}
	\label{tab:rad_obs}
	\begin{tabular*}{\textwidth}{m{25mm} M{16mm} M{16mm} M{15mm} M{15mm} m{18mm} c}
		\hline
		\hline
		Source & Project ID & \shortstack{Date\\(2015)} & Integration time (min) & Array configuration & Observing band (GHz)& Phase calibrator \\
		\hline \\[-8pt]
		Cen\,X--4 & 14B-117 & 17 Jan & 34 & CnB & X (8.0--12.0) & J1522-2730\\ \\[-7pt]
		IGR\,J17511--3057 & 14B-153 & 26 Mar & 13 & B & X & J1744--3116\\
		                  &         & 2 Apr  & 13 &  & X & \\
		                  &         & 8 Apr & 12 &  & X & \\ \\[-7pt]
		SAX\,J1808.4--3658 & 14B-153 & 18 Apr & 11 & B & X & J1744--3116\\
		                   &         & 19 Apr & 11 &  & K (18.0--26.5)& \\
		                   &         & 19 Apr & 15 &  & C (4.0--8.0)& \\
		                   &         & 22 Apr & 11 &  & X & \\
		                   &         & 27 Apr & 12 &  & X & \\
		                   &         & 5 May  & 11 &  & X & \\
		                   &         & 7 May  & 22 &  & X & \\
		                   &         & 10 May & 23 &  & X & \\ \\[-7pt]
		IGR\,J00291+5934 & 15B-339 & 11 Aug & 36 & A & X & J0102+5824\\
	                    &         & 15 Aug & 36 &  & X & \\
	                   &         & 18 Aug & 36 &  & X & \\ \\[-7pt]
		\hline
	\end{tabular*}
	\end{minipage}
\end{table*}

\subsection{Data analysis}

\subsubsection{VLA}

The radio observations of all four systems are summarised in Table~\ref{tab:rad_obs}. We calibrated the radio data from the VLA with the Common Astronomy Software Applications package \citep[CASA v4.2.1;][]{2007ASPC..376..127M} using standard procedures. The source 3C48 was used as a flux calibrator for IGR\,J00291+5934, and 3C286 for the other sources. For all four sources, we switched between phase calibrator and science target on 5 minute cycles (3 minutes for the 22\,GHz observations). We imaged the Stokes I data using Briggs weighting with a robustness parameter of 1 as a compromise between sidelobe contamination and point source sensitivity. We performed deconvolution using the multi-frequency, multi-scale \texttt{clean} algorithm with two Taylor coefficients to account for sky frequency dependence. No self-calibration was performed. The flux density of each detected source was measured by fitting the source in the image plane using the \texttt{imfit} task. We forced a point-source fit, with an elliptical Gaussian of the shape of the synthesised beam. The flux uncertainty is taken as the rms of the residual image, with an additional systematic uncertainty of 2\%, typical of flux calibration with the VLA at these frequencies. Given that no self-calibration was performed, the flux errors could be underestimated. The observing frequencies (primarily centred at 10\,GHz), and the short phase-referencing cycle times (5 min), should help minimize the effects of phase decorrelation. While the lack of other point sources in the fields of view of our targets means we cannot quantify the level of phase decorrelation, we assume it to be negligible for the reasons stated above. Our final flux density measurements are given in Table~\ref{tab:fluxes}. The flux density upper limits on non-detections are reported at 3 times the local rms.

To measure the radio spectral index of SAX\,J1808.4--3658, we split each 10~GHz observation into two sub-bands (8--10\,GHz and 10--12\,GHz) and imaged them separately. On April 19, we carried out 6~GHz and 22~GHz observations for a more precise measurement of the spectral index. We use the flux densities measured at each of these two frequencies to obtain the spectral index $\alpha$ following the $S_{\nu} \propto {\nu}^{\alpha}$ convention (where $S_{\nu}$ is the flux density measured at the central frequency $\nu$).

We also calibrated previously unpublished VLA observations of IGR J00291+5934 taken during the 2004 (program AR545 on December 11 and 14, taken in A configuration at 4.86 GHz) and 2008 (program MPRTST on September 30 in D configuration at 8.46 GHz) outbursts. As in our 2015 observations, the phase calibrator was J0102+5824 for all these observations, and the flux calibrator was 3C286 in 2004 and 3C147 in 2008. For these archival observations, we used only one Taylor coefficient during deconvolution (suitable for these single-frequency observations) and natural weighting (to maximize point source sensitivity).

\subsubsection{Swift}

\textit{Swift}/XRT data were analysed with \textsc{HEAsoft}, as part of the \textsc{HEASARC} software suite \citep{1996ASPC..101...17A}. To generate the \textit{Swift}/BAT light curves, we used the \textit{Swift}/BAT light curve database \citep{2013ApJS..209...14K}.

During our single X-ray observation of Cen\,X--4 on January 17, 2015 (MJD 56650.4), \textit{Swift}/XRT operated in photon-counting mode (PC, 2.5\,s resolution). Within a $20^{\prime\prime}$ radius of its position, only 7 total counts are detected (corresponding to a count rate of $7.5 \times 10^{-3}$\,counts/s). The background level was measured in an area offset from visible X-ray sources. Given the low number of counts, we do not fit its spectrum, but assume the same spectral shape as that reported by \citet{2013MNRAS.436.2465B} for the faintest state of Cen\,X--4 (count rate $<0.07$\,counts/s), and use Poisson statistics \citep{1991ApJ...374..344K} to estimate the 90\% confidence interval for its count rate. We assume a low-luminosity spectrum for Cen\,X--4 consisting of a thermal component (which dominates at energies below $\approx 2$\,keV) and a power-law component ($\Gamma = 2.0$), absorbed by a column density $N_{\rm H} = 8 \times 10^{20}$\,cm$^{-2}$ \citep{2013MNRAS.436.2465B}. When calculating the uncertainty in the count rate, we only take into consideration the errors associated with the low number of counts, since they will dominate over the errors associated with the spectral model. We use XSPEC (12.9.0) and WebPIMMS\footnote{https://heasarc.gsfc.nasa.gov/docs/software/tools/pimms.html} \citep{1993Legac...3...21M} to convert from count rate to unabsorbed flux.

For IGR\,J17511-3057, we use the light curve and spectral information provided by \citet{2016A&A...596A..71P}, observed with \textit{Swift}/XRT in window-timing (WT, 1.76\,ms resolution) and PC modes. They fit the spectra with an absorbed power-law, and reported the fluxes in the 0.5--10\,keV range (for the same 6.9\,kpc distance assumed in this work), which we converted to 1--10\,keV fluxes in WebPIMMS, using their measured absorption column and photon indices.

For SAX\,J1808.4--3658, we analyzed all pointed Swift/XRT observations taken between April 13, 2015 (MJD 57125.4) and May 19, 2015 (MJD 57161.8). A total of 26 observations were taken, with Program IDs 33034, 33737 and 81453. The X-Ray Telescope operated either in WT or PC mode. The source counts were extracted in an energy range of 1--10 keV, as is commonly used in recent literature, from a circular region of radius $20^{\prime\prime}$, and the background was extracted from an area far from known sources in the field. The events and X-ray spectra were extracted with the tool XSELECT (v2.4d) and the spectral analysis was done with XSPEC. Each spectrum was fitted with a simple absorbed power-law model plus a multiplicative component (\textsc{cflux}) used to calculate the unabsorbed flux in the 1--10\,keV band (and its 90\% confidence interval).

For IGR\,J00291+5937, we use a procedure similar to that described for SAX\,J1808.4--3658. The observation campaign carried out with \textit{Swift}/XRT on this source will be described in full in Russell et al., in preparation.

\subsubsection{Ariel~V}
For the 1979 outburst of \mbox{Cen\,X--4}, we use the X-ray data reported by \citet{1980ApJ...241..779K}, observed with the Ariel~V All-Sky X-ray Monitor \citep{1976Ap&SS..42..123H}, and retrieved from the public archive. We converted the reported 3--6\,keV flux to 1--10\,keV unabsorbed flux using WebPIMMS, by assuming a power-law spectrum with photon index $\Gamma = 1.6$ (as adopted by \citealp{2016ApJ...826..149B}) and absorption ($N_{\rm H} = 8 \times 10^{20}$\,cm$^{-2}$; \citealp{2013MNRAS.436.2465B}).

\subsubsection{RXTE}
To estimate the X-ray fluxes of IGR\,J00291+5934 at those times when the source was not detected in the radio band during its outbursts in 2004 and 2008, we used the X-ray data reported by \citet{2005ApJ...622L..45G} and \citet{2010A&A...517A..72L}, observed with the RXTE Proportional Counter Array
(PCA). These data were retrieved from the XTE Mission-Long Source Catalog. We converted the 2--9\,keV count rate from IGR\,J00291+5934 to the 1--10\,keV unabsorbed flux using WebPIMMS, by assuming a power-law spectrum with photon index $\Gamma = 2.06$ and absorption $N_{\rm H} = 4.3 \times 10^{21}$\,cm$^{-2}$, which describes the 0.4--10\,keV spectrum \citep{2005A&A...444..357P}.

\subsubsection{LCO}
The $i^{\prime}$-band magnitudes were extracted using \textsc{phot} in IRAF and calibrated using the three comparison stars listed in \cite{Greenhill06}. $i^{\prime}$-band magnitudes of the three comparison stars were calculated from $R$-band and $I$-band magnitudes using the conversion of \cite{Jordi06}, in the same way as mentioned in \cite{Elebert09}.

\section{Results}

\subsection{Cen\,X--4}

We did not detect \mbox{Cen\,X--4} in the radio band, although we place constraints on the emission (<14\,$\mu$Jy; Table~\ref{tab:rad_obs}) that are a factor of $\approx 20$ deeper than previous radio observations in quiescence \citep{1992xbfb.work...99K}. In the X-ray band, we detected only 7 total photons (including background) from \mbox{Cen\,X--4}, as our observations were conducted during a particularly faint X-ray state (Fig.~\ref{fig:cen_x-4_lc}). The 90\% confidence interval for its background-subtracted 1--10\,keV luminosity is $L_{\rm X} = 8.0 \times 10^{30}$ -- $5.6 \times 10^{31}$\,erg\,s$^{-1}$.

\subsection{IGR\,J17511--3057}

We conducted radio observations of IGR\,J17511--3057 during its slow X-ray decay (Fig.~\ref{fig:j17511_lc}). Similar to previous outbursts, it remained undetected in the radio band, although our constraints (<21\,$\mu$Jy; Table~\ref{tab:rad_obs}) are a factor of $\approx 8$ deeper than previous radio observations at similar X-ray luminosity \citep{2009ATel.2232....1M}.

\subsection{IGR\,J00291+5934}

Our radio observations were conducted during the fast X-ray decay of IGR\,J00291+5934 (Fig.~\ref{fig:j00291_lc}), and we detected it in the first two of three epochs (Table~\ref{tab:rad_obs}). We did not detect the source in the archival observations taken on December 11 (<132\,$\mu$Jy) and 14 (<111\,$\mu$Jy), 2004 or on September 30, 2008 (<156\,$\mu$Jy). Coupled with previous detections, IGR\,J00291+5934 now has a well-sampled X-ray luminosity range, meaning that we can track its disk--jet coupling from the peak of the outburst towards quiescence (Fig.~\ref{fig:lrlx_3sources}), assuming that it follows the same path over multiple outbursts, as seen in the black holes GX\,339--4 \citep{2013MNRAS.428.2500C} and V404 Cyg \citep{2017ApJ...834..104P}.

\subsection{SAX\,J1808.4--3658}

\subsubsection{Light curves} \label{sec:lc1808}

The seven radio detections of SAX\,J1808.4--3658 (labelled sequentially A--G) at different times across a single outburst make it an excellent source for tracing the inflow--outflow coupling in an AMXP. Observations during previous outbursts have identified SAX\,J1808.4--3658 as a radio-loud neutron star \citep{2006MNRAS.366...79M}, and in addition to the main outburst, we are now also able to track its multiwavelength behaviour in the fainter states (Fig.~\ref{fig:j1808_lc}).

In X-rays, SAX\,J1808.4--3658 behaved during its 2015 outburst similarly to previous outbursts: a slow decay (until MJD $\approx 57141$), a fast decay (MJD $\approx 57141 - 57143$), and a reflaring tail (after MJD $\approx 57143$). A radio reflare was also observed, but the X-ray and radio reflares do not seem to be coincident (observations F, G in Fig.~\ref{fig:j1808_lc}). Radio detections during X-ray reflares show that the jet survives during the reflaring tail. The last two radio observations show unexpected behaviour, as the source undergoes a radio reflare during an X-ray minimum (observation F, although an X-ray reflare at that time could have been missed, see Section~\ref{sec:sparse_lc}), and no increase in radio flux is observed during a rapid increase in X-ray flux at the start of an X-ray reflare (observation G). We find little evidence of intra-observation variability during these two radio observations (consistent with a constant flux over 2\,min and 10\,min timescales within $\approx 2 \sigma$ uncertainty).

\begin{figure}
	\begin{minipage}{\columnwidth}
		\includegraphics[width=.99\textwidth]{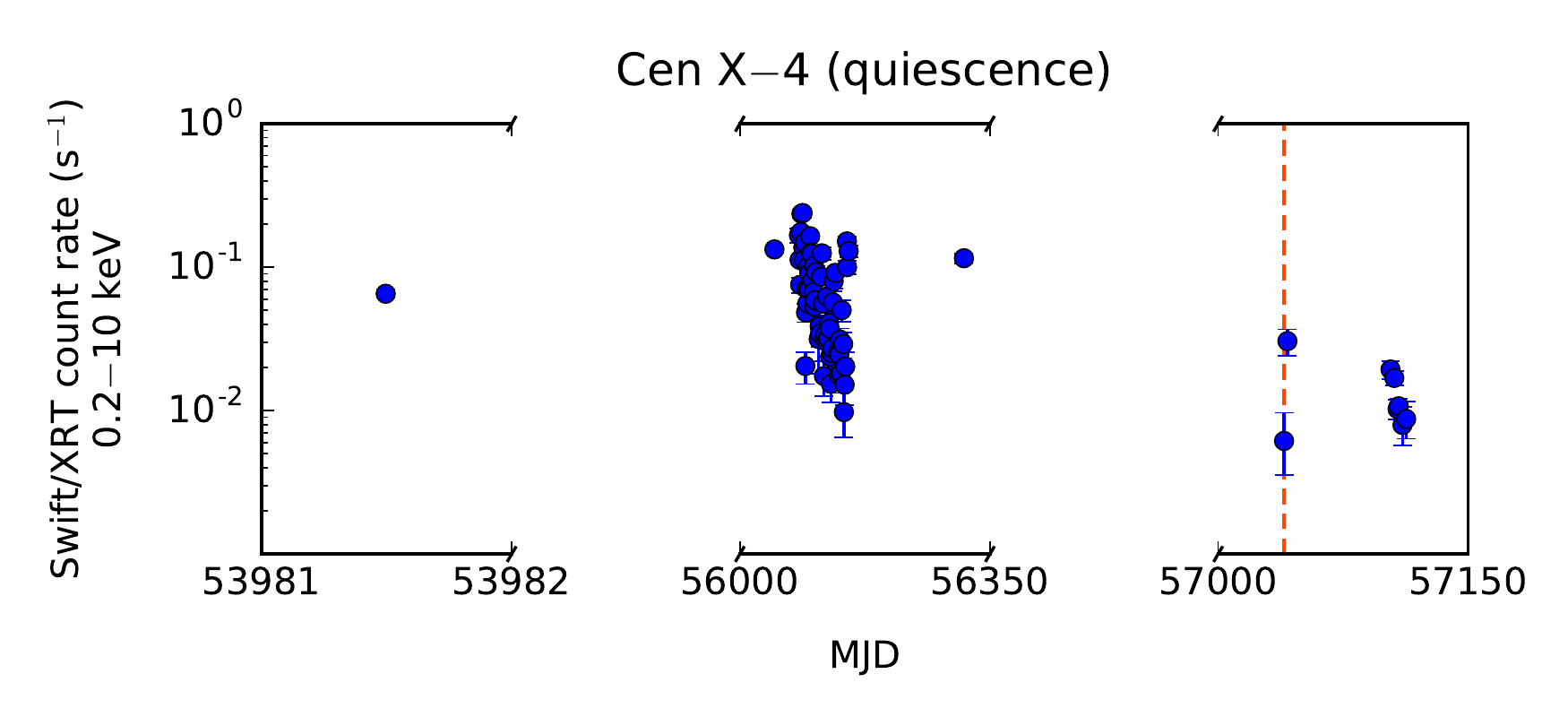}
	\end{minipage}
	\begin{minipage}{\columnwidth}
		\includegraphics[width=.99\textwidth]{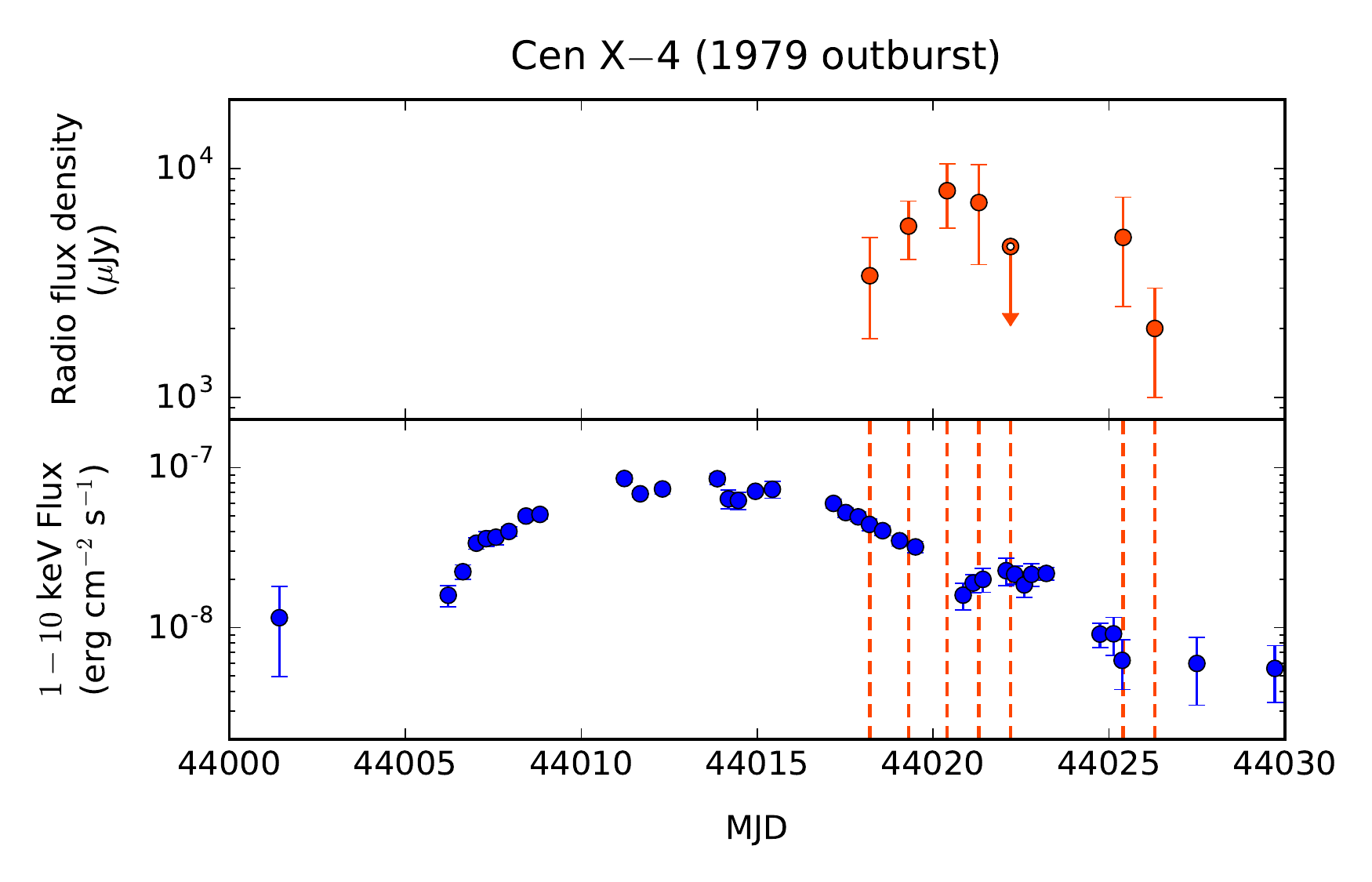}
	\end{minipage}
	\caption{\textit{Top}: quiescent X-ray light curve of \mbox{Cen\,X--4}, as observed by \textit{Swift}/XRT in PC mode. We show all observations with the \textit{Swift}/XRT (over the years 2006/2012/2015) as a visual representation of the quiescent flux of Cen\,X--4. The blue points show individual snapshots and the vertical dashed line indicates the timing of our VLA observations in 2015, carried out simultaneously with an XRT observation when the system was in a particularly faint state. \textit{Bottom}: The radio and X-ray light curves of \mbox{Cen\,X--4} during its 1979 outburst, as observed with the VLA \citep{1988ApJ...335L..75H} and the Ariel~V satellite \citep{1980ApJ...241..779K}. \mbox{Cen\,X--4} was detected in the radio band multiple times during the X-ray decay.}
	\label{fig:cen_x-4_lc}
\end{figure}

\begin{figure}
	\includegraphics[width=.99\columnwidth]{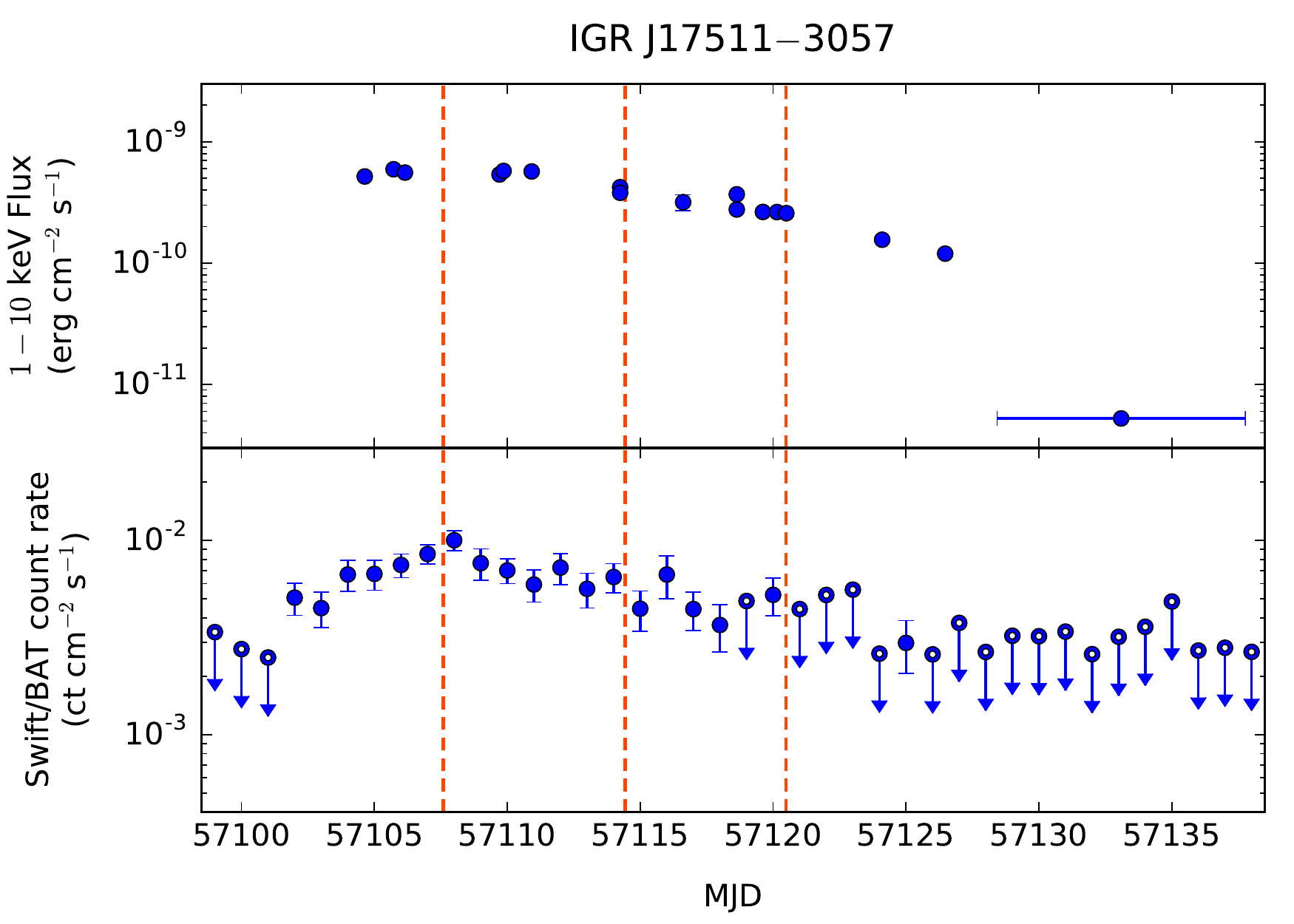}
	\caption{X-ray light curves of IGR\,J17511--3057 during its outburst in 2015, as observed by \textit{Swift}/XRT (1--10\,keV flux) and \textit{Swift}/BAT (15--150\,keV count rate). The blue points show individual observations and the vertical dashed lines indicate the timing of our VLA observations. All radio observations were carried out during the slow X-ray decay. We did not detect the source in any of our VLA observations.}
	\label{fig:j17511_lc}
\end{figure}

\begin{figure}
	\centering
	\includegraphics[width=.99\columnwidth]{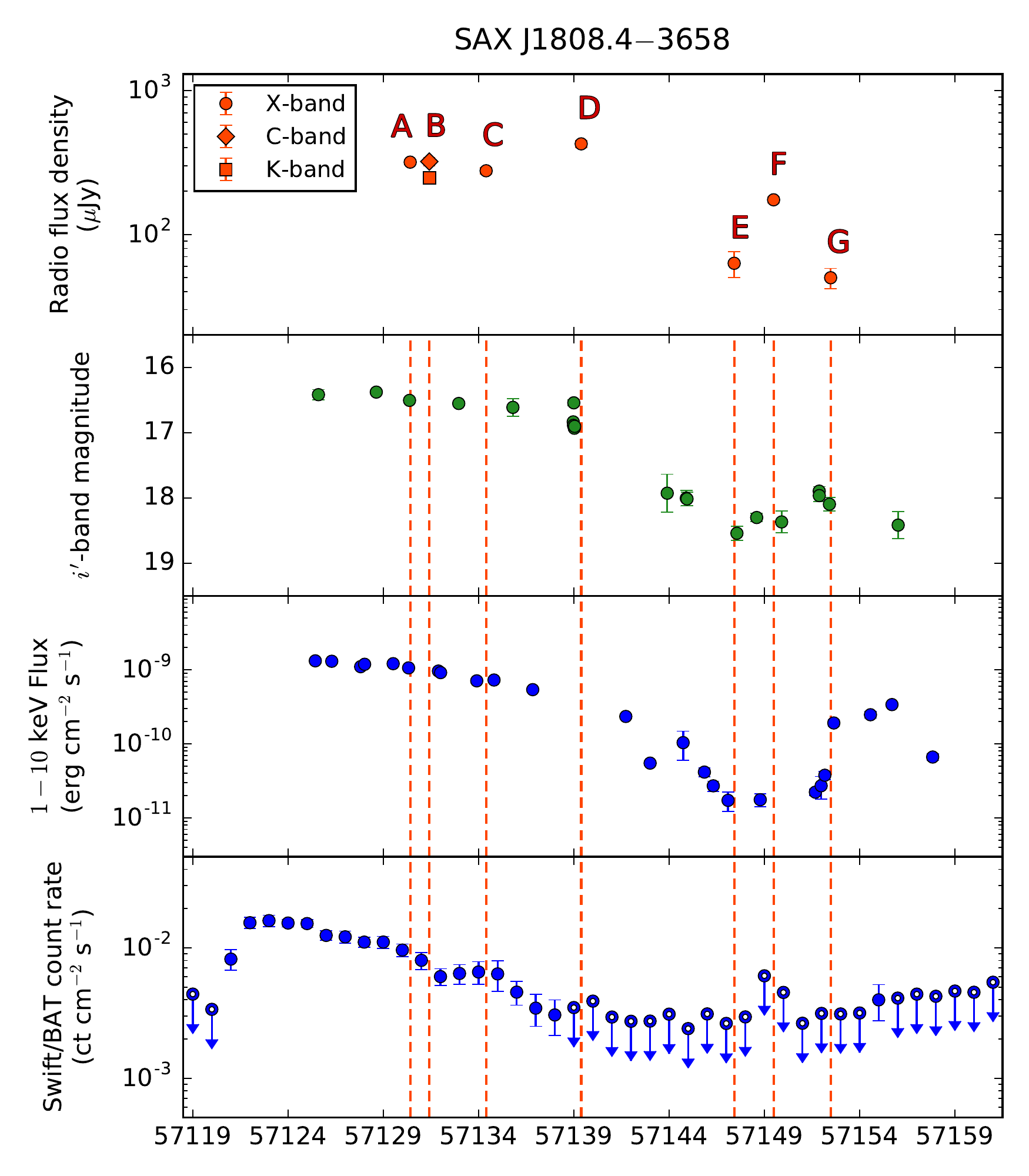}
	\caption{Radio (VLA), optical (LCO) and X-ray (\textit{Swift}/XRT, 1--10\,keV flux; \textit{Swift}/BAT, 15--150\,keV count rate) light curves of J1808.4--3658 during its 2015 outburst. The orange dashed lines indicate the times of the radio observations, individually labelled from A to G. During the slow X-ray decay, the source undergoes a radio reflare (observation F).}
	\label{fig:j1808_lc}
\end{figure}

\begin{figure}
	\centering
	\includegraphics[width=.99\columnwidth]{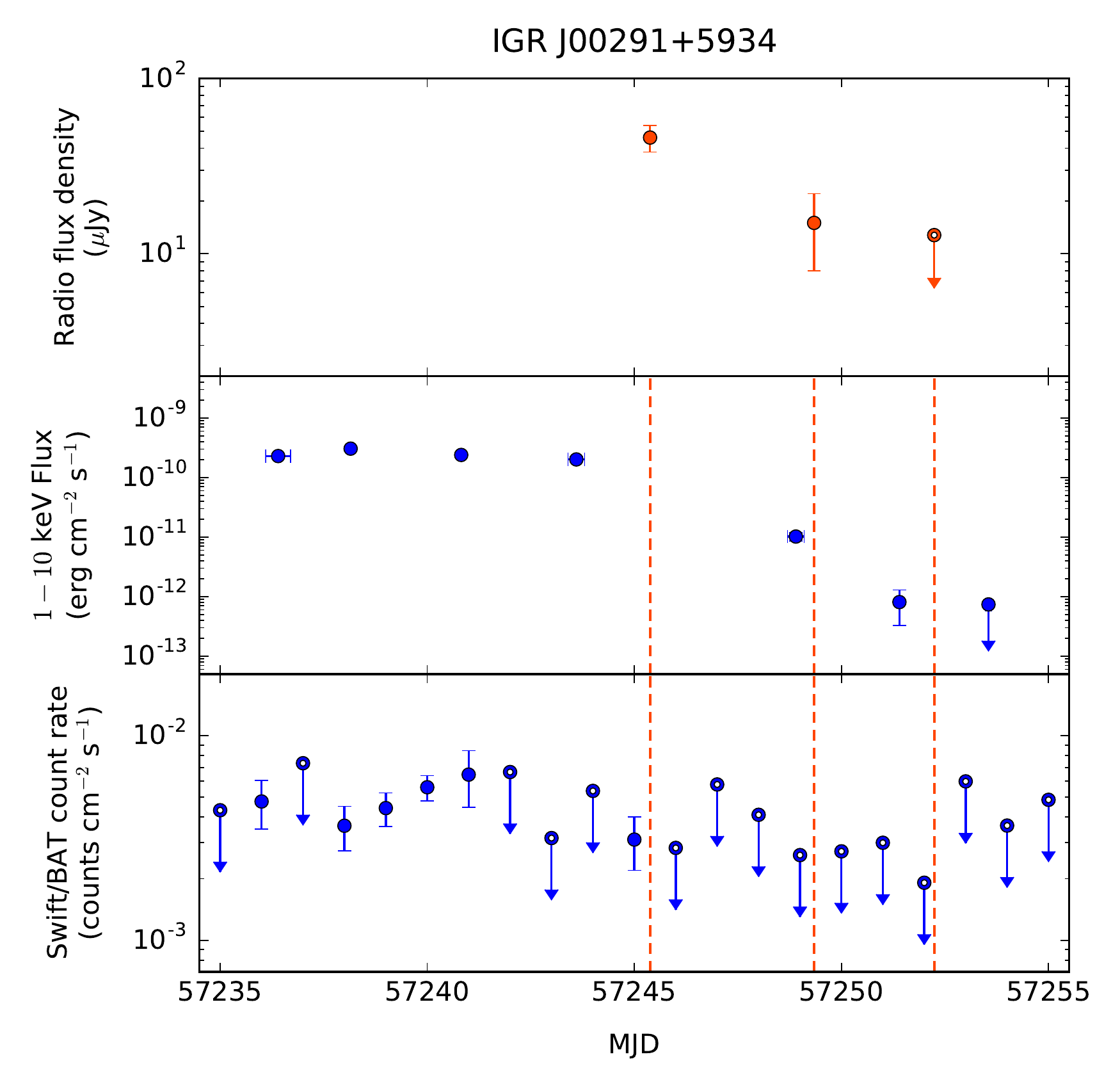}
	\caption{The radio (VLA) and X-ray (\textit{Swift}/XRT, 1--10\,keV flux; \textit{Swift}/BAT, 15--150\,keV count rate) light curves of IGR\,J00291+5934 during the late stages of its 2015 outburst. We detect the source in the first two out of our three observations.}
	\label{fig:j00291_lc}
\end{figure}

\subsubsection{Radio spectral indices} \label{sec:spec_ind}

The 6 and 22\,GHz observations from April 19 were taken within 20 minutes of each other, during which time we do not expect the source to have varied (see \ref{sec:lc1808}). Indeed, we do not find evidence for variability during either observation on that day (both C and K band data are consistent with a constant flux within each observation within $1\sigma$ uncertainty). We find the spectral index to be close to flat ($\alpha = -0.24 \pm 0.10$, assuming no change in flux between the two observations), similar to the compact jets of other X-ray binaries, although we cannot rule out a steep spectrum within uncertainties. We find that the spectral index measurements within the 10\,GHz band on the other observing epochs are typically unconstraining (with a typical error of $\pm 0.5$, due to the faintness of the radio emission and the small lever arm in frequency), so we cannot distinguish between optically thin and thick emission. The spectral index measurements for all epochs are listed in Table~\ref{tab:fluxes}.

\begin{table*}
	\centering
	\begin{minipage}[b]{0.88\textwidth}
	\caption{Radio monitoring results. Source distances are assumed as discussed in Section \ref{sec:targets}.}
	\label{tab:fluxes}
	\begin{tabular*}{\textwidth}{m{25mm} M{15mm} M{15mm} M{25mm} M{15mm} M{15mm} l}
		\hline
		\hline \\[-8pt]
		Source &  
		\shortstack{D \\ (kpc)} & MJD &
		$F_{\rm 1-10\,keV}^{\rm a}$ ($\times 10^{-11}$\,erg\,cm$^{-2}$\,s$^{-1}$) &
		\shortstack{$\nu$ \\ (GHz)} &
		\shortstack{$S_\nu$ \\ ($\mu$Jy\,bm$^{-1}$)} &
		\multicolumn{1}{c}{$\alpha$} \\
		\hline \\[-8pt] 
		Cen\,X--4 & $ 1.2 $ & 56650.409 & $(8.5 \pm 5.4) \times 10^{-3}$ & 10.0 & $<14$ &\multicolumn{1}{c}{--\hphantom{\,$\rm ^c$}} \\ \\[-7pt]
		IGR\,J17511--3057 & $ 6.9 $ & 57107.592 & $58.5\pm10.1$ & 10.0 & $<22$ & \multicolumn{1}{c}{--\hphantom{\,$\rm ^c$}} \\
						  &				& 57114.432 & $38.9\pm6.7$ & 10.0 & $<21$ & \multicolumn{1}{c}{--\hphantom{\,$\rm ^c$}} \\
						  &				& 57120.494 & $27.0\pm4.6$ & 10.0 & $<28$ & \multicolumn{1}{c}{--\hphantom{\,$\rm ^c$}} \\ \\[-7pt]
		SAX\,J1808.4--3658 & $ 3.5 $ & 57130.411 & $98.7\pm5.5$ & 10.0 & $317 \pm 24$ & \hphantom{$-$}$0.05 \pm 0.45$\\								& & 57131.413 & $91.4\pm5.1$ & 22.0 & $ 238 \pm 23$ & \multicolumn{1}{c}{--\,$\rm ^b$} \\
								& & 57131.427 & $91.4\pm5.1$ & 6.0 & $321 \pm 25$ & $-0.24 \pm 0.10$\,$\rm ^b$\\
								& & 57134.397 & $73.0\pm4.1$ & 10.0 & $277 \pm 22$ & $-0.51 \pm 0.56$~\\
								& & 57139.386 & $50.0\pm2.8$ & 10.0 & $426 \pm 24$ & $-0.21 \pm 0.60$\\
								& & 57147.415 & $1.6\pm0.1$ & 10.0 & $63 \pm 13$ & \multicolumn{1}{c}{--\,$\rm ^c$}\\
								& & 57149.484 & $1.9\pm0.1$ & 10.0 & $174 \pm 9$ & $-0.26 \pm 0.57$\\
								& & 57152.478 & $11.1\pm0.6$ & 10.0 & $50 \pm 8$ & \multicolumn{1}{c}{--\,$\rm ^c$}\\ \\[-7pt]
								
		IGR\,J00291+5934 & 4.2 & 57245.376 & $19.6\pm7.2$ & 10.0 & $46 \pm 8$ & \multicolumn{1}{c}{--\,$\rm ^c$}\\
			& & 57249.335 & $0.65\pm0.16$ & 10.0 & $15 \pm 7$ & \multicolumn{1}{c}{--\,$\rm ^c$}\\
			& & 57252.241 & $(3.6\pm2.8) \times10^{-2}$& 10.0 & $<12$ & \multicolumn{1}{c}{--\hphantom{\,$\rm ^c$}}\\
									  
		\hline \\[-8pt]
	\end{tabular*}
	\textbf{Notes.}
	
	$\rm ^a$\,Estimated at the time of the radio observation
	
	$\rm ^b$\,The 6 and 22 GHz observations were used to derive a single spectral index measurement.
	
	$\rm ^c$\,The signal-to-noise ratio is too low for a meaningful spectral index measurement.
	\end{minipage}
\end{table*}

\subsection{The radio--X-ray correlation}

\begin{figure}
	\centering
	\includegraphics[width=.99\columnwidth]{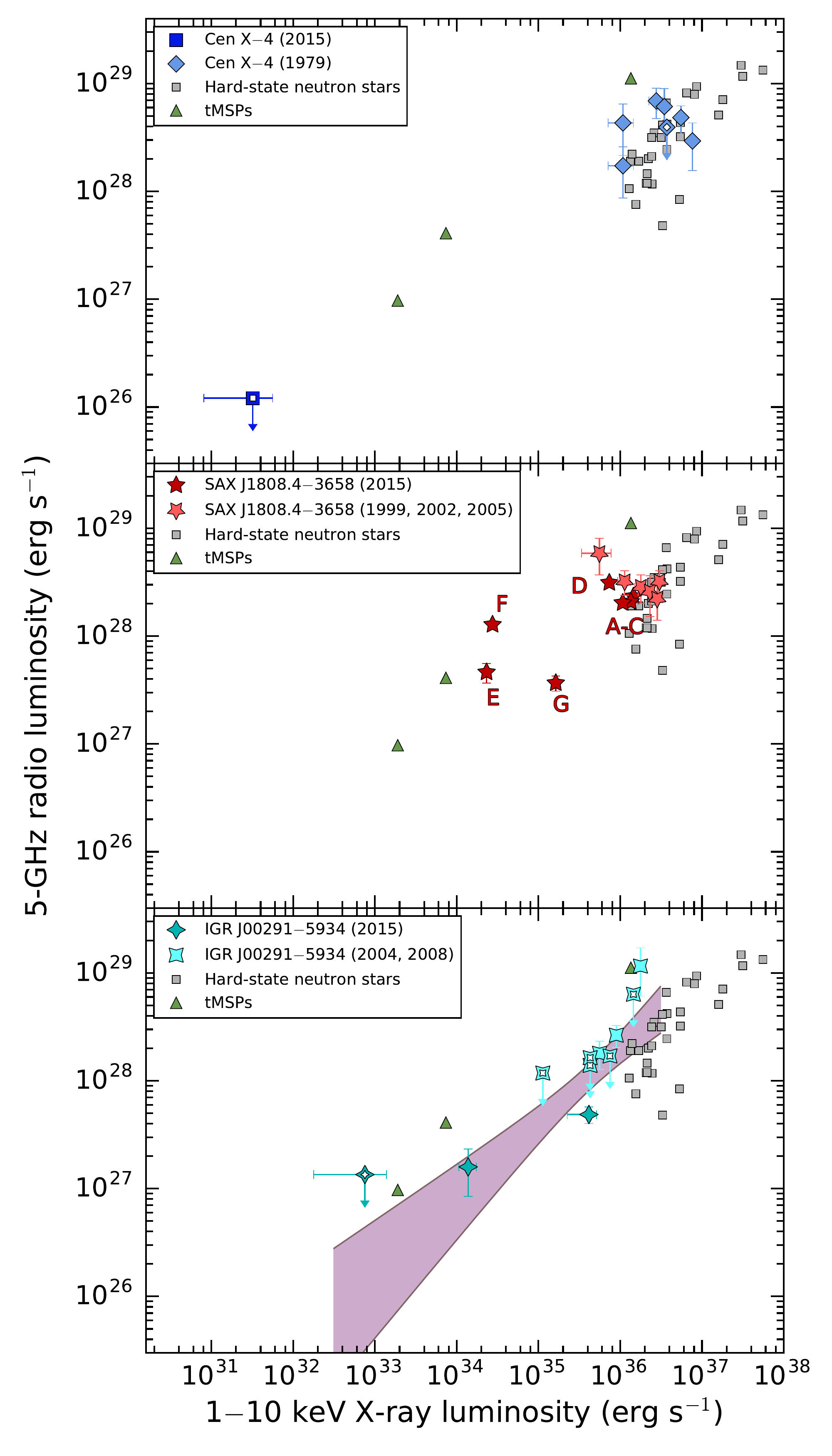}
	\caption{The radio and X-ray luminosities of \mbox{Cen\,X--4}, SAX\,J1808.4--3658 and IGR\,J00291+5934. Hollow symbols mark upper limits. The labelled data points of SAX\,J1808.4--3658 refer to the same observations in Fig.~\ref{fig:j1808_lc}. IGR\,J00291+5934 (and possibly SAX\,J1808.4--3658, but with larger scatter) seems to follow the familiar $L_{\rm R} \propto L_{\rm X}^{\beta}$ correlation seen in black holes and some atoll sources, with $\beta = 0.77 \pm 0.18$ (90\% confidence interval, shaded area).}
	\label{fig:lrlx_3sources}
\end{figure}

\begin{figure*}
	\centering
	\includegraphics[width=.95\textwidth]{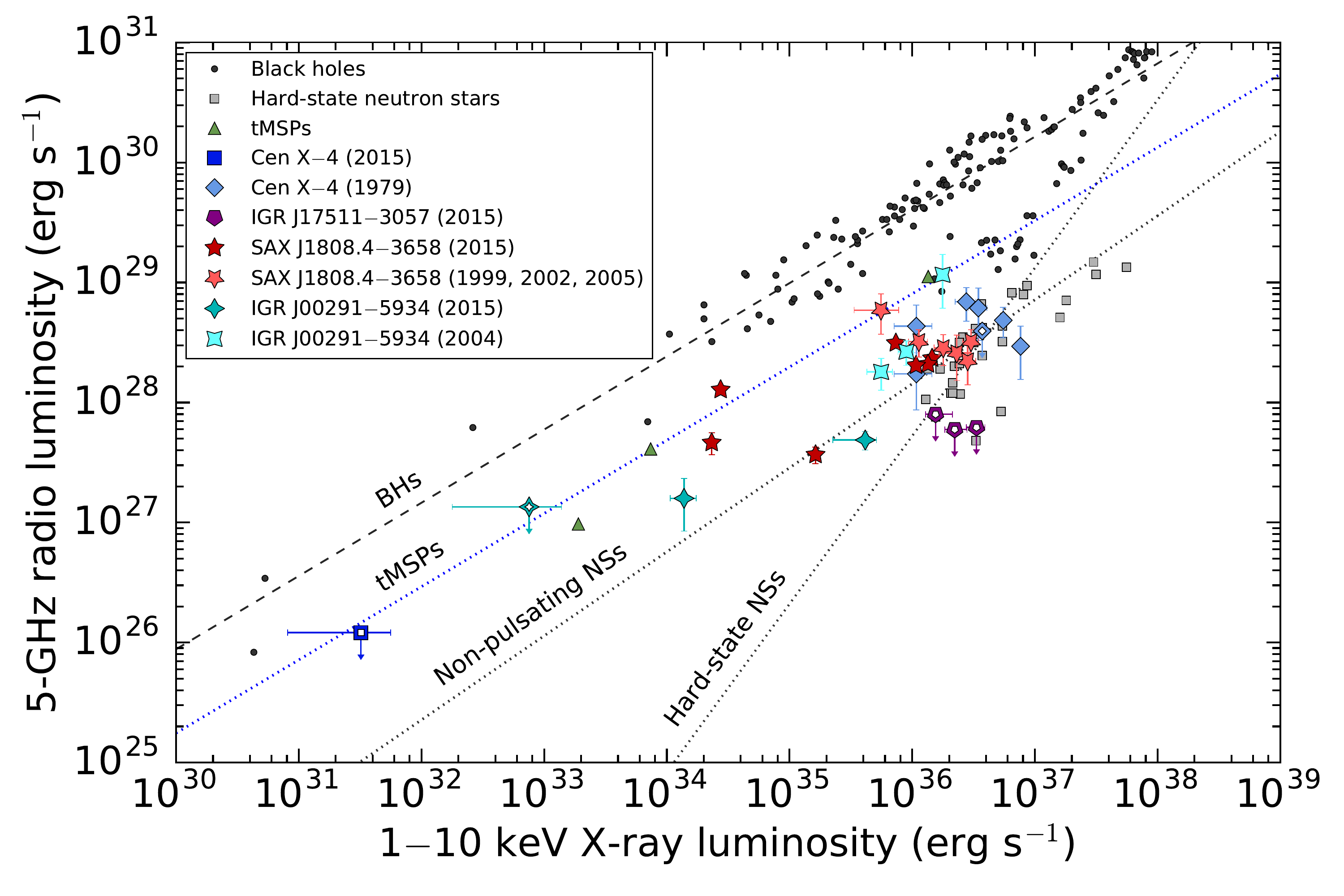}
	\caption{Radio and X-ray luminosities of low-mass X-ray binaries. The black dashed line represents the radio--X-ray correlation of accreting black holes \citep[$\beta=0.6$;][]{2014MNRAS.445..290G}, and the three dotted lines are the proposed relations for tMSPs \citep[$\beta=0.6$;][]{2015ApJ...809...13D}, non-pulsating neutron stars (NS) regardless of spectral state ($\beta=0.7$), and hard-state neutron stars only \citep[$\beta=1.4$;][]{2006MNRAS.366...79M, 2011MNRAS.415.2407M}. Hollow symbols indicate upper limits. The upper limits from the previous outbursts of IGR\,J00291+5934 are not plotted for better visibility.}
	\label{fig:lrlx}
\end{figure*}

The radio ($L_{\rm R} = 4 \pi D^2 \nu S_{\nu}$, where $D$ is the distance to the source) and X-ray luminosities of our sample are displayed in Fig.~\ref{fig:lrlx_3sources} individually, and together in Fig.~\ref{fig:lrlx}, with the addition of other black hole and neutron star systems from the literature.

Positions of X-ray binaries on the radio\,--\,X-ray plane are best measured from strictly simultaneous observations. Among our multi-wavelength campaign, \mbox{Cen\,X--4} has strictly simultaneous data. For the other sources, we interpolate the X-ray light curve to estimate the X-ray flux at the time of the radio observations, by fitting a piecewise linear function in the log\,$L_{\rm X}$/time space.

The 3\,$\sigma$ upper limit of 14\,$\mu$Jy on the radio emission of \mbox{Cen\,X--4} in quiescence only constrains its radio luminosity to being less than that of the radio-fainter black hole jets (e.g. XTE\,J1118+480; \citealp{2014MNRAS.445..290G}). The limit lies just above the proposed tMSP track, slightly below the quiescent black hole systems. 

In addition, we also place the 1979 outburst of \mbox{Cen\,X--4} on the radio/X-ray plane for the first time. For the radio luminosity, we use the data reported by \citet{1988ApJ...335L..75H} in two bands (at 1.49 and 4.9\,GHz). For days when observations were carried out in both radio bands, we use the measurement with smaller uncertainties, with the assumption of a flat spectrum. We find that during the 1979 outburst, \mbox{Cen\,X--4} reached radio and X-ray luminosities similar to the atoll sources. Assuming it follows a single, continuous radio--X-ray correlation from the peak of the outburst down to quiescence (like the black hole systems V404~Cyg, \citealp{2008MNRAS.389.1697C}; and GX\,339--4, \citealp{2013MNRAS.428.2500C}), we obtain a lower limit for its correlation slope $\beta \gtrsim 0.5$.

The upper limits for IGR\,J17511--3057 make it less luminous in the radio band than typical atoll sources (Fig.~\ref{fig:lrlx}), but we cannot rule out radio emission at a level similar to that of EXO\,1745--248, the radio-detected neutron star with the lowest disk--jet coupling normalization \citep{2016MNRAS.tmp..785T}.

Previous radio detections of SAX\,J1808.4--3658 sample just one decade of X-ray luminosity \citep[$L_{\rm X} \approx 10^{35}$ to $10^{36}$\,erg\,s\,$^{-1}$;][]{2011MNRAS.415.2407M}. Here, we extend the range of detections down to $L_{\rm X} \approx 10^{34}$\,erg\,s\,$^{-1}$. The radio behaviour of SAX\,J1808.4--3658 during the slow X-ray decay seems to be reproducible between different outbursts, with a clustering of detections at $L_{\rm X} \approx 10^{36}$\,erg\,s\,$^{-1}$ and $L_{\rm R} \approx 10^{28}$\,erg\,s\,$^{-1}$. As it fades, SAX\,J1808.4--3658 is seen to alternate between the atoll sources and tMSPs in the radio--X-ray plane. This points to a temporal changes in the radiative efficiency, which does not reach the same levels as for black hole systems. 

Similar to SAX\,J1808.4--3658, our radio observations of IGR\,J00291+5934 make it possible to study the jet over two decades of X-ray luminosity, in the range $L_{\rm X} \approx 10^{34}$ to $10^{36}$\,erg\,s\,$^{-1}$. In contrast to SAX\,J1808.4--3658, IGR\,J00291+5934 is consistent with following a $L_{\rm R} \propto L_{\rm X}^\beta$ correlation. We find a correlation slope $\beta = 0.77 \pm 0.18$ (90\% uncertainty range, assuming reproducible outbursts), which is similar to the primary black hole track, to the proposed tMSP track, and to Aql\,X--1 (above $10^{36}$\,erg\,s\,$^{-1}$; \citealp{2016MNRAS.tmp..785T}), at a luminosity similar to that of atoll sources.

\section{Discussion}

Our results reinforce the idea of different behaviours in different neutron star systems -- with reflares and no consistent radio--X-ray relationship (SAX\,J1808.4--3658), with a well-fitting radiatively inefficient flow (IGR\,J00291+5934), long-term steady, but with flipping between X-ray modes (tMSPs), etc. Our observations strengthen the evidence for atoll sources and AMXPs being fainter in the radio band than black holes at the same X-ray luminosity. Some AMXPs can be faint (IGR\,J00291+5934) or fainter than non-pulsating neutron stars (IGR\,J17511--3057). During outbursts, however, some AMXPs (SAX\,J1808.4--3658) can become as radio-bright as tMSPs (a conclusion also supported by the tMSP M28I, which was observed in outburst as an AMXP; \citealp{2013Natur.501..517P}). 

The above hints at a maximum radio luminosity achievable by neutron stars, a factor of $\approx$ 5 fainter than radiatively inefficient black holes at the same X-ray luminosity, as proposed by \citet{2015ApJ...809...13D}. One expects radio jets from more massive accretors to be more luminous, as indicated by the fundamental plane of accreting black holes ($\log L_{\rm R} \propto 0.8 \log M$; \citealp{2003MNRAS.345.1057M, 2004A&A...414..895F}). Therefore, as already noted by \citet{2015ApJ...809...13D}, the difference in mass between black holes ($\approx 8 M_\odot$; \citealp{2010ApJ...725.1918O, 2012ApJ...757...36K}) and neutron stars ($\approx 1.4M_\odot$; \citealp{2012ApJ...757...55O}) could account for a factor of $\approx$ 4 between the radio luminosity of the two classes of X-ray binaries, if their accretion mechanisms are otherwise similar. This could possibly explain some of the difference between the radiatively inefficient black holes, and the most radio-luminous neutron stars (the tMSPs, and the radio-loud epochs of SAX\,J1808.4--3658).

Because of different disk--jet coupling behaviours in different neutron stars, highlighted by the contrast between IGR\,J00291--3658, with a coupling index   $\beta \approx 0.7$, and EXO\,1745--248, with $\beta \approx 1.7$ \citep{2016MNRAS.tmp..785T}, we do not attempt to fit a single radio/X-ray correlation for the whole population of neutron stars (as previously done by \citealp{2006MNRAS.366...79M}). Such a fit would be biased towards radio-loud neutron stars, and would thus not offer further insights into the accretion process, nor better predictive power. As a tool to identify new black holes and neutron star candidates below $L_{\rm X} \lesssim 10^{36}$\,erg\,s$^{-1}$, we therefore advise only using the maximal radio luminosity (neutron stars at least a factor of $\approx$ 5 fainter than black holes) suggested above.

Below we discuss each source in our sample, considering in most detail the case of SAX\,J1808.4--3658, the source for which we obtained the best-sampled multi-wavelength coverage.

\subsection{Cen\,X--4}

The X-ray flux of \mbox{Cen\,X--4} during our observations was an order of magnitude lower than its median quiescent flux (see Fig.~\ref{fig:cen_x-4_lc}). Its quiescent X-ray luminosity \citep{2010ApJ...720.1325C, 2013MNRAS.436.2465B}, is typically in the range $L_{\rm X} = 0.7-4.3 \times 10^{32}$\,erg\,s$^{-1}$ in the 1--10\,keV band (assuming a photon index $\Gamma$ = 1.5). In contrast, we measure a luminosity $L_{\rm X}$ $\approx 0.05 - 0.24 \times 10^{32}$\,erg\,s$^{-1}$ (90\% uncertainty range), which means that we cannot place constraints on jet production in \mbox{Cen\,X--4} in quiescence, deeper than that of some black holes.

During its outburst in 1979, Cen\,X--4 had similar radio ($L_{\rm R} = 10^{28} - 10^{29}$\,erg\,s$^{-1}$) and X-ray luminosities ($L_{\rm X} = 10^{36} - 10^{37}$\,erg\,s$^{-1}$) as most of the non-pulsating systems. The signal-to-noise ratio of these early radio detections, however, is too low to check if the two wavebands are correlated. Including our radio upper limit during quiescence, the radio--X-ray correlation index is only restricted to $\beta \gtrsim 0.5$, so we cannot differentiate between radiatively efficient or inefficient accretion in this system.

Had we observed \mbox{Cen\,X--4} at a more typical quiescent X-ray luminosity, a radio observation would have likely either detected or disproved the formation of a jet as powerful as that of tMSPs. We therefore encourage future simultaneous radio and X-ray observations of \mbox{Cen\,X--4}.

\subsection{IGR\,J17511--3057}

The radio non-detection of IGR\,J17511--3057 could potentially be attributed to several factors: an incorrect distance estimate, a quenching of the jet in the soft state, or suppression of jet formation by a high magnetic field. IGR\,J17511--3057 would need to be located approximately 30\,kpc away for its upper limits to be consistent with the $\beta = 0.7$ correlation for neutron stars. Given that multiple distance estimates point towards a source distance of less than 7\,kpc (see Section \ref{sec:intr_j17511}), we consider that its non-detection has a physical origin. Jet quenching at high Eddington ratios has previously been observed in the atoll sources \mbox{Aql\,X-1} (\mbox{\citealp{2009MNRAS.400.2111T}}, \citealp{2010ApJ...716L.109M}), GX\,9+9 \citep{2011IAUS..275..233M}, and 4U\,1728--34 \citep{2003MNRAS.342L..67M}, but it has not been seen in all neutron stars \citep{1998ATel....8....1R, 2004MNRAS.351..186M}. The hard spectral state and power spectrum of the low luminosity outburst of IGR\,J17511--3057 \citep{2016A&A...596A..71P} point towards a canonical hard state \citep{2006ApJ...643.1098S} and hence, against the non-detections being caused by jet quenching in the soft state.

The high magnetic fields of pulsars may also inhibit the formation of jets \citep[for fields $\gtrsim 10^{11}$\,G;][]{2000MNRAS.317....1F, 2012IJMPS...8..108M}, although the magnetic field of IGR\,J17511--3057 has been estimated to be similar to that of other AMXPs ($\sim 10^8$\,G; \citealp{2015MNRAS.452.3994M}, \mbox{\citealp{2016A&A...596A..71P}}). The detection of type\,I bursts \citep{2010MNRAS.409.1136A} also rules out a high magnetic field \citep{1980ApJ...238..287J}.

Given that radio emission at the same level as EXO\,1745--248 is not ruled out, it is possible that IGR\,J17511--3057 has an X-ray to radio normalization that is lower than that of other AMXPs and atolls. Contrary to the common assumption that AMXPs are more radio-loud than non-pulsating sources, IGR\,J17511--3057 would then be a clear example that this rule does not hold for all AMXPs. Since IGR\,J17511--3057 has a similar spin ($\nu = 245$\,Hz) and magnetic field strength ($B \approx 10^8$\,G) to other AMXPs \citep{2015MNRAS.452.3994M}, it appears that the magnitude of the spin and magnetic field are not the only properties that produce strong jets in the other AMXPs and tMSPs, and that other factors, as yet unidentified, could be at least as important. 

\subsection{SAX\,J1808.4--3658}

The atypical radio behaviour of SAX\,J1808.4--3658 might be interpreted in a variety of ways, which we explore below. Throughout this section, we refer to our radio epochs as A--G, as labelled in Fig.~\ref{fig:j1808_lc}. Of these, we refer to those epochs that are close to the proposed radio/X-ray correlation for tMSPs as ``radio-loud'' (our observations D, E, F, and the detection by \citealp{1999ApJ...522L.117G}).

\subsubsection{Insufficient light curve sampling}
\label{sec:sparse_lc}
Sparse X-ray coverage could be invoked to explain the apparent high radio luminosity of the radio-loud epochs, and their deviation from the proposed radio/X-ray tracks for neutron stars. For example, \textit{Swift}/XRT observations might have missed an X-ray reflare coincident with the radio reflare of observation F. The X-ray reflares can last as little as $\approx$ 0.3~days \citep{2001ApJ...560..892W}. In comparison, the two \textit{Swift}/XRT observations adjacent to the radio reflare (observation F) are separated by $\approx$ 3~days. We therefore examine hard X-ray (\textit{Swift}/BAT, 15--50\,keV) and optical (LCO) light curves for evidence of flares or dips of other wavebands at the times of the radio observations. We find the optical and soft X-ray fluxes in SAX\,J1808.4--3558 track each other well during the outburst (Fig.~\ref{fig:j1808_lc}; a full description of the optical--X-ray correlation will be presented in Bernardini et al.), similar to previous outbursts \citep{2016ApJ...817..100P}. In previous outbursts, however, the X-ray lagged the optical emission by 1.5--4 days \citep{1999MNRAS.304...47G, 2016ApJ...817..100P}, and the optical and X-ray fluxes during some reflares were anti-correlated or uncorrelated \citep{2006tpr..conf...53W, 2016ApJ...817..100P}. Given the lack of \textit{Swift}/XRT data, we therefore look for evidence of reflares primarily in the \textit{Swift}/BAT light curve.

For radio observations D, E, F, the optical and hard X-ray light curves do not show evidence of reflares (with the previously described caveat of possible uncorrelated X-ray/optical behaviour). In addition, observation E was probably carried out at the end of a stable X-ray decay (as shown by the preceding \textit{Swift}/XRT observations, MJD $\sim$57144 to $\sim$57148). In observation F, the radio flux density increased by a factor of $\approx$ 3 over the two days since observation E, with no apparent corresponding increase in the 1--10\,keV flux. Unfortunately, the coincidentally large uncertainties in the daily \textit{Swift}/BAT data adjacent to this radio observation were not particularly constraining ($F_{\rm 1-10\,keV} < 10^{-10}$\,erg\,cm$^{-2}$\,s$^{-1}$, $3\sigma$ limit, assuming a photon index $\Gamma=1.8$), so a relatively faint ($L_{\rm X} < 1.5 \times 10^{35}$\,erg\,s$^{-1}$), fast X-ray flare without a corresponding optical flare could have been missed. There was also no significant detection in the orbital \textit{Swift}/BAT light curve around observation F. We find it unlikely that the other radio-loud epochs were caused by insufficient sampling of its multi-wavelength light curve.

Observations D, E, F, together with previous radio observations, show that SAX\,J1808.4--3658 can reach quasi-simultaneous radio/X-ray luminosities similar to those of tMSPs. The sparse 1--10\,keV sampling around observations D and F could potentially be the reason behind their apparently high radio luminosities, but observation E and that of \citet{1999ApJ...522L.117G}, might require a physical explanation. Below, we investigate possible physical causes.

\subsubsection{A collision of the jet with a nearby medium}
\label{sec:colli}

The enhanced flux during the radio-loud epochs could be caused by additional radio emission from a shock. Thus, a significant fraction of the radio emission could originate from radio lobes or hot spots from the interaction of the jet with the interstellar medium. We do not resolve any extended emission in SAX\,J1808.4--3658 that might indicate lobes. Based on the size of the synthesised beam at 22\,GHz ($3.4^{\prime\prime} \times 0.9^{\prime\prime}$), we place an upper limit of $\lesssim 2 \times 10^{17}$\,cm on the size of the radio source (at a distance of 3.5\,kpc). Separations between the two radio lobes produced by outbursting X-ray binaries have been seen to reach similar dimensions \citep[e.g.][]{2005ApJ...633..218B}. Travelling the above distance at velocity {\it c} takes 0.2 years, however. Since this is longer than the outburst timescale, the reflares cannot be associated with lobes of such size. We cannot dismiss the possibility of smaller radio lobes based on physical size alone. However, the low density environment of SAX\,J1808.4--3658 \citep[$N_{\rm H} = 1.4 \times 10^{21}$\,cm$^{-2}$, inferred from the X-ray spectrum;][]{2013A&A...551A..25P} argues against the presence of a dense interstellar medium in its vicinity for a jet to interact with. However, we still cannot rule out the (somewhat unlikely) possibility of a cloud adjacent to the neutron star that does not intervene along our line of sight towards it, which requires a finely-tuned geometry.

Instead of the high-density interstellar medium that SAX\,J1808.4--3658 would need to be embedded in, the jet could interact with a smaller, localized circumbinary structure, produced by accretion/ejection processes in the system, from the current or previous outbursts. Such material would likely lead to accelerated orbital decay \citep{2001ApJ...561..329T}. SAX\,J1808.4--3658, however, undergoes orbital expansion \citep{2012ApJ...746L..27P}, which we take as evidence against the existence of a massive circumbinary structure around the system.

Alternatively, the radio emission could be caused by collisions of a high-velocity collimated jet with slow, wide-angle winds. This scenario is thought to occur in SS433, a black hole system accreting close to the Eddington luminosity, which produces heavy outflows driven by strong radiation pressure \citep{2011ApJ...735L...7B}. The coexistence of jets and winds is thought to occur in a few other black hole and neutron star systems at high X-ray luminosities \citep[$L_{\rm X} \gtrsim 0.3 \: L_{\rm Edd}$;][]{2016ApJ...830L...5H}. In contrast, SAX\,J1808.4--3658 only reaches luminosities $L_{\rm X} \lesssim 0.1 \: L_{\rm Edd}$, although it does launch winds \citep{2014A&A...563A.115P}. If winds are still launched down to $L_{\rm X} = 10^{33}$\,erg\,s$^{-1}$ (possibly by the propeller mode of accretion; \citealp{2009MNRAS.399.1802R, 2014MNRAS.441...86L}), the jet could blow into them, producing the observed radio--X-ray behaviour. In comparison, compact radio and $\gamma$-ray emission from cataclysmic variables in outburst originates at the interface between slow and fast outflows \citep{2014Natur.514..339C}. The possibility of coexisting jets and winds in SAX\,J1808.4--3658 can be tested in future outbursts with simultaneous radio and high spectral resolution X-ray observations.

Another possibility is that the radio flares are produced by a jet colliding with the accretion disk or the donor star. This could occur if the spin of the neutron star is severely misaligned with respect to the orbit of the donor star, and the accretion disk is warped \citep{1975ApJ...195L..65B, 2002MNRAS.336.1371M, 2003ApJ...587..748B}. If this were the case, we should expect the same mechanism to operate in some black hole systems. The fact that no black hole low-mass X-ray binary in the hard state is known to display as much scatter in the radio/X-ray plane as SAX\,J1808.4--3658, suggests that such a geometry and emission mechanism are unlikely. In addition, the small donor \citep[$M = 0.03-0.06 M_\odot$;][]{2013ApJ...765..151W}, which covers $\approx$ 1\% of the sky as seen from the neutron star, makes a collision between it and the jet unlikely.

Information about the nature of the radio emission can in principle be gleaned from the spectral index. Radio lobes typically have steep spectral indices ($\alpha \approx -0.7$), similar to discrete, transient jets, whereas steady, compact jets have flat spectra \citep[$\alpha \approx 0$; ][]{1978BAAS...10..629B, 2001MNRAS.322...31F}. As discussed in section \ref{sec:spec_ind}, the best constrained spectral index measurement of SAX\,J1808.4--3658 is $\alpha = -0.24 \pm 0.10$, but we cannot confidently (<3$\sigma$ confidence) favour any emission mechanism, although a flat-spectrum jet during the slow decay is slightly more favourable. The slightly shallower spectral index than expected for neutron stars (with an average $\alpha \approx 0.2$; \citealp{2007MNRAS.379.1108R}) could imply contamination from optically thin regions.

Another observational test that would confirm collisions of the jet with a surrounding medium would be a rebrightening in the radio band simultaneous with an increase in optical line flux (e.g. H$\alpha$), during constant, or decaying, optical continuum level \citep[e.g.][]{2005Natur.436..819G}.

\subsubsection{Collisions within the jet}

Shocks within a jet have previously been invoked to explain the radio properties of jets in general \citep{2000A&A...356..975K, 2004MNRAS.355.1105F, 2010MNRAS.401..394J, 2013MNRAS.429L..20M} and the behaviour of individual X-ray binaries in particular \citep{2001ApJ...553L..27F, 2012MNRAS.423.2656R}. A luminosity-dependent Lorentz factor of the ejecta (invoked by \citealp{2011MNRAS.413.2269S} and \citealp{2015MNRAS.450.1745R} to explain the X-ray/radio luminosity tracks of radio-faint black holes) could in principle lead to delayed radio reflares. Fast ejecta launched at the peak of an X-ray reflare could collide with previous slower ejecta, and cause a radio brightening of the source. The typical durations and peak fluxes of the reflares of SAX\,J1808.4--3658, which are so far unique to this source, might be responsible for the scatter in the radio--X-ray plane that has not been observed in other X-ray binaries.

\subsubsection{A large delay between the radio and X-ray emission}
The scatter in the radio/X-ray plane could be caused by the rapid variability of SAX\,J1808.4--3658 and a large delay between the radio and X-ray light curves. In the black hole X-ray binary GX\,339--4, infrared emission from the jet lags X-ray emission by 100\,ms \citep{2010MNRAS.404L..21C}; in another black hole system, GRS~1915+105, X-rays lead the radio by $\approx$ 0.1\,h \citep{1998A&A...330L...9M}. Such a short time-lag ($\ll 1$\,h) cannot explain the radio-loud observations (D, E, F) in SAX\,J1808.4--3658. A neutron star with a different ejection mechanism (e.g. the propeller mode of accretion/ejection) could potentially show a longer delay between the radio and X-ray bands (on the order of hours). Such a time delay produced by the propeller effect could be common among AMXPs, but it has so far not been detected due to the poor radio sampling of AMXP outbursts. 

Still, delays of less than a day would only give a significantly different result for the last radio observation (G), such that its corresponding X-ray flux would be an order of magnitude fainter, making it radio-loud.

It could also be that the radio emission is delayed with respect to the X-ray emission by a few days. Given the sparse sampling of the radio light curve, a cross-correlation test for finding the delay between the radio and X-ray emission will not be robust for our data set. Below, we highlight two possible time delays, but a detailed quantitative analysis will require better sampled multi-wavelength light curves.

It is plausible that the radio flare of observation F could actually be produced by the X-ray reflare at MJD $\sim$57144, with a delay of $\sim$5~days. Such long delays have previously been observed between the core and knots in the jet of the neutron star X-ray binary Cir\,X--1 \citep{2004Natur.427..222F}. The knots in the jets of Cir\,X--1, however, are resolved on 1$^{\prime\prime}$ scales, and are likely to be produced during an interaction with the surroundings, which we do not find evidence for in SAX\,J1808.4--3658 (see Section \ref{sec:colli}). 

Less likely (as outflows typically follow inflows) is for the radio emission to precede the X-ray emission such that observation F and the reflare at MJD $\gtrsim$~57152 are directly linked. However, a possible explanation for the radio preceding the X-ray emission could be a timescale for the diffusion through the magnetosphere that is longer than the ejection timescale, such that a fraction of an infalling supply of gas is ejected before its remainder penetrates the magnetosphere, as seen in the simulations of \citet{2014MNRAS.441...86L}.

\subsubsection{The ejection of matter in the ``propeller'' mode}
The radio emission could come from material ejected by the propeller effect, as suggested by \citet{1999ApJ...522L.117G}, who detected the strongest radio emission from SAX\,J1808.4--3658 a day after the start of the fast X-ray decay during its outburst in 1998. In the propeller mode of accretion, the magnetic pressure of the magnetosphere balances the ram pressure of the infalling matter, accelerating the inner regions of the accretion disk \citep{1975A&A....39..185I}. If the gas velocity exceeds the escape velocity of the neutron star, outflows are launched from the system (strong propeller). Such outflows prevent most of the disk material from accreting onto the surface of the neutron star, decreasing the radiative efficiency of the flow \citep{2000ApJ...541..849C}. However, matter may accumulate in the disk, and accrete quasi-periodically on the neutron star even in the strong propeller mode \citep{2014MNRAS.441...86L}.
 
In the case of SAX\,J1808.4--3658, the propeller effect has been invoked as a possible mechanism behind the rapid X-ray decay during its outbursts \citep{1998A&A...338L..83G}, the re-flares following the main outburst \citep{2009ApJ...707.1296P, 2016ApJ...817..100P}, and the light curve modulations seen during the main outburst \citep[1--5\,Hz;][]{2014ApJ...789...99B} and reflares \citep[1\,Hz;][]{2009ApJ...707.1296P}. This makes it a possible candidate to explain the variable radio emission. At low luminosities, the jet would be dominated by propeller emission, and at higher luminosities it would transition to a jet similar to those of atoll sources. The dipolar and stable magnetic field of SAX\,J1808.4--3658 \citep{1999PhRvL..83.3776L} could play a part in the efficient ejection of matter at low accretion rates.

We develop a toy model to explore whether the strong propeller mode of accretion could account for the radio-loud epochs of SAX\,J1808.4--3658. Given the radio and X-ray properties of tMSPs in the accreting state, in which the propeller is thought to operate \citep{2015ApJ...809...13D}, we expect the kinetic and emission properties of propeller-driven jets to be similar to those of other X-ray binary jets (collimated, flat spectrum). The radio luminosity might therefore be expected to scale with the kinetic power \citep{2003MNRAS.343L..59H} according to:
\begin{equation}
\label{eq:lr}
L_{\rm radio} = L_{\rm 0} \left(\frac{W_{\rm jet}}{W_{\rm 0}}\right)^{1.42}\ .
\end{equation}
From approximations of the kinetic and radiative powers of the jets of three radio galaxies, \citet{2005ApJ...633..384H} found $W_0 = 6.2 \times 10^{37}$\,erg\,s$^{-1}$, with an estimated order of magnitude uncertainty, by setting $L_{\rm 0} = 1.6 \times 10^{30}$\,erg\,s$^{-1}$, as determined empirically from the fundamental plane of accreting black holes \citep{2003MNRAS.345.1057M, 2004A&A...414..895F}. The large uncertainties come from the difficulty of measuring the kinetic power of jets, and from the orders of magnitude difference between the radiative and kinetic energies of jets.

To estimate the kinetic energy of an outflow, we use energy conservation in the system, which is expressed as:
\begin{equation}
\label{eq:etot}
\dot{E}_{\rm pot} + \dot{E}_{\rm sd}= L_{\star} + L_{\rm flow} + W_{\rm jet}\ ,
\end{equation}
where $\dot{E}_{\rm pot}$ is the rate of gravitational potential release, $\dot{E_{\rm sd}}$ is the spin-down power supplied by the neutron star, $L_{\star}$ and $L_{\rm flow}$ are the radiated luminosities from the impact stream onto the surface of the neutron star and in the accretion flow, and $W_{\rm jet}$ is the kinetic energy carried by the jet. The total rate of gravitational potential energy release is given by:
\begin{equation} \label{eq:epot}
\dot{E}_{\rm pot} = \frac{GM\dot{M}_d}{R_{\star}}\ ,
\end{equation}
where $M$ and $R_{\star}$ are the mass and radius of the neutron star respectively.

To estimate the kinetic power flowing into the jet, we use the analysis of \citet[Fig.~2]{2015MNRAS.449.2803D} for a 2\,ms accreting pulsar with a $10^8$\,G field, which was based on the numerical work of \citet{2006ApJ...646..304U} and \citet{2014MNRAS.441...86L}. Their formulation for the radiative efficiency ($\mathcal{F} = (L_{\rm flow} + L_{\star}) / \dot{E}_{\rm pot}$) shows what fraction of the gravitational potential energy is radiated. When the accretion disk truncates below the corotation radius, all matter accretes onto the neutron star, resulting in a radiative efficiency $\mathcal{F} = 1$. At larger truncation radii (set by the magnetospheric radius), matter is redirected into jets, decreasing $\mathcal{F}$ below unity. We can estimate the radiative power of the system as:
\begin{equation}
\mathcal{F} \dot{E}_{\rm pot} = GM\dot{M}_d \left( \frac{1 - f_{\rm out}}{R_{\star}} + \frac{1}{2} \frac{f_{\rm out}}{R_{\rm in}} \right)
\end{equation}
(where the factor of 1/2 is from the virial theorem), which we use to estimate the fraction of mass lost to the jet ($f_{\rm out}$). This assumes that the fraction of radiative power originating from spin-down is negligible. Indeed, we expect less than 1\% of the spin-down power to be converted to X-rays \citep{2002A&A...387..993P}. To obtain a lower limit for the kinetic power of the jet ($W_{\rm jet} = 0.5 f_{\rm out} \dot{M}_d v_{\rm jet}^2$), we assume matter is ejected at the escape velocity ($v_{\rm jet} = \sqrt{2GM/R_{\rm in}}$).

In the propeller regime, the jet draws power from the spin of the neutron star, which in turn loses rotational energy at a rate $\dot{E}_{\rm sd} = 4 \pi^2 I \nu \dot{\nu}$, where $I$ is the moment of inertia of the neutron star ($I = 10^{45}$\,g\,cm$^2$). We provide an upper limit on the power of the jet by assuming a constant spin-down rate of $\dot{\nu} = 2.5 \times 10^{-14}$\,Hz\,s$^{-1}$ (the upper limit found by \citealp{2008ApJ...675.1468H} during outbursts), and that the value of $W_{\rm 0}$ is one order of magnitude lower than reported by \citet{2005ApJ...633..384H} (who estimate an order of magnitude uncertainty on $W_{\rm 0}$). When $R_{\rm in}$ is close to the corotation radius, it is expressed as \citep{1993ApJ...402..593S}:
\begin{equation}
\label{eq:rin}
R_{\rm in} \approx \left( \frac{\mu^2}{4 \Omega \dot{M}_d} \right)^{1/5}\ ,
\end{equation}
where $\mu$ is the magnetic moment of the star ($\mu = B R_{\star}^2$, where $B$ is the magnetic field strength).

\begin{figure}
	\includegraphics[width=0.99\columnwidth]{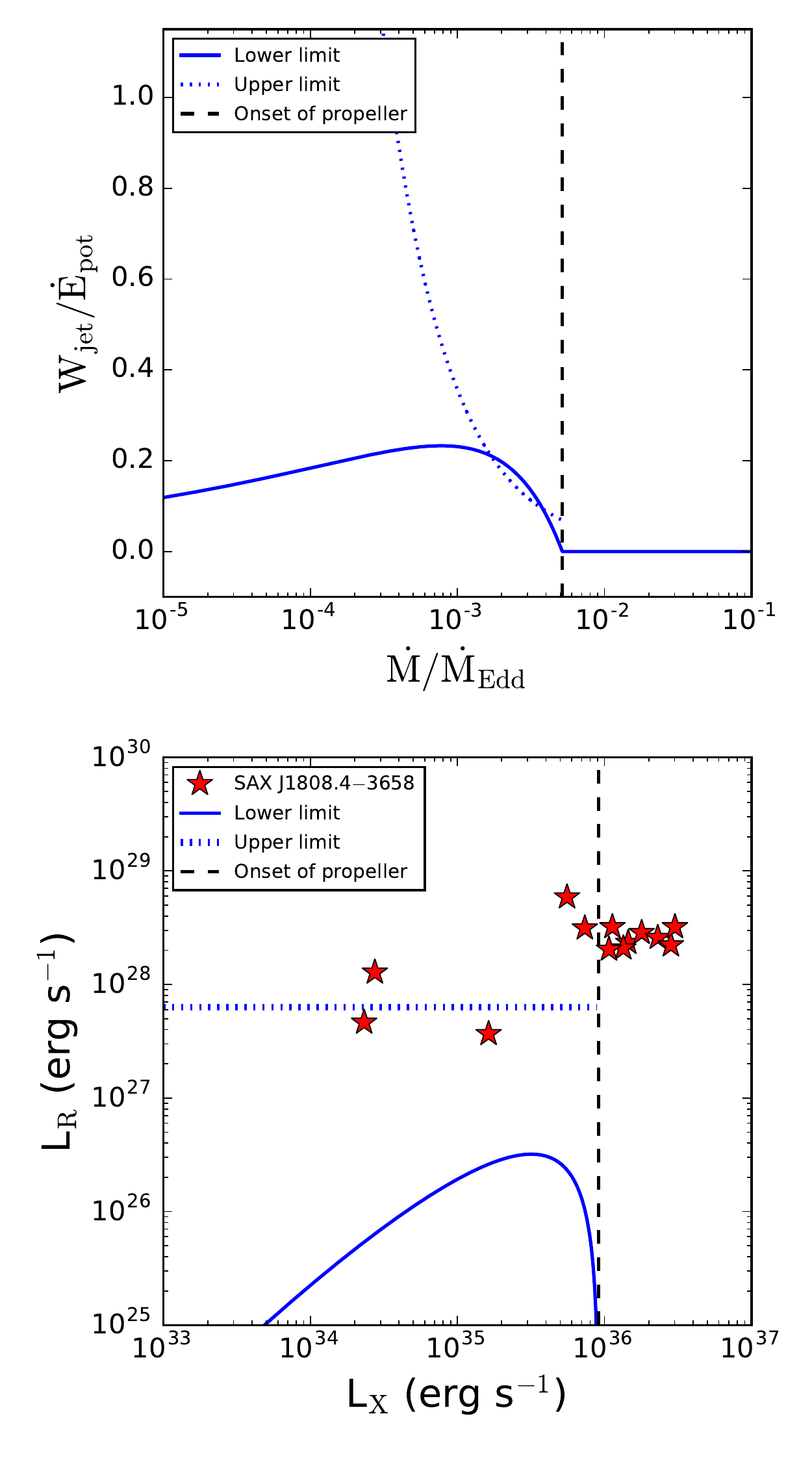}
	\caption{\textit{Top}: Ratio between jet power and gravitational potential energy release, given as a lower limit (continuous line) and upper limit (dotted line). \textit{Bottom}: Radio emission from the strong propeller for our toy model, using the analysis of \citealp{2015MNRAS.449.2803D} (as a lower limit), and using the upper limit on spin-down energy (as an upper limit). The jet is of SAX\,J1808.4--3658 is unlikely to be caused by a propeller effect described by our model.}
	\label{fig:propmod}
\end{figure}

The X-ray luminosity in the 1--10\,keV band represents a fraction of the radiative power of the system, such that  $L_{\rm X} = \eta \: \mathcal{F} \: \dot{E}_{\rm pot}$ ($\eta = 0.5$ is the bolometric correction), and the radio luminosity is given by Equation \ref{eq:lr}. Under these assumptions, the inner disk radius is 2.5 times the corotation radius at $L_{\rm X}=10^{33}$\,erg\,s$^{-1}$, making Equation~\ref{eq:rin} valid at the X-ray luminosities sampled here ($L_{\rm X}=10^{34}-10^{37}$\,erg\,s$^{-1}$). Fig.~\ref{fig:propmod} (bottom) shows the results of our model. We find that the radio luminosity estimate from the spin-down of the neutron star is compatible with the values derived using the analysis of \citet{2015MNRAS.449.2803D}. The empirical upper limit on the spin-down rate during the outbursts of SAX\,J1808.4--3658 \citep[$\dot{\nu} < 2.5 \times 10^{-14}$\,Hz\,s$^{-1}$;][]{2008ApJ...675.1468H} is also compatible with the maximum spin-down rate in the above model ($\dot{\nu} = 1.7 \times 10^{-14}$\,Hz\,s$^{-1}$). We find, however, that the radio luminosity predicted by our model is at least one order of magnitude too low for it to account for the behaviour of SAX\,J1808.4--3658, making it unlikely that the propeller effect is solely responsible for its radio-bright jet. This discrepancy might be explained by a large fraction of the gravitational potential energy being directed to the jet, or by the poorly known scaling between the power of the jet and radio luminosity, especially in the case of the propeller regime. In addition, simulations show that two kinds of outflows are driven as a consequence of the strong propeller: a slow, matter-dominated disk wind, and a collimated magnetically-dominated Poynting jet \citep{2006ApJ...646..304U, 2009MNRAS.399.1802R, 2014MNRAS.441...86L}. In contrast, our model assumes a jet typical of X-ray binaries, which are often considered to be energetically equipartitioned (not magnetically dominated). Moreover, some X-ray emission is likely to originate at the magnetosphere--disk interface, and the flow is thought to be less radiatively efficient \citep{2015ApJ...807...33P}. A full theoretical model that takes into account all these effects and an evolving fraction of ejected mass is beyond the scope of this paper.

To provide more rigorous tests of the connection between propeller outflows and radio luminosity, additional simultaneous radio and X-ray observations of propeller-mode sources are necessary. Focusing on luminosities at which the strong propeller is thought to switch on will assist in quantifying the differences between standard X-ray binary jets and propeller jets. To robustly determine whether propeller outflows are responsible for strong radio emission, an empirical characterisation of the connection between the kinetic and radiative signatures of propeller-mode systems is required. Hydrodynamical simulations would be needed to fully explain the observed emission in tMSPs, the radio-loud and radio-faint AMXPs.

\subsection{IGR\,J00291+5934}

For the first time, we report a radio--X-ray correlation index $\beta$ for a neutron star low-mass X-ray binary in the range $L_{\rm X} \approx 10^{34}$ to $10^{36}$\,erg\,s\,$^{-1}$, using the detections of IGR\,J00291+5934. In contrast to SAX\,J1808.4--3658, IGR\,J00291+5934 shows correlated decay in the two bands. A correlation slope $\beta = 0.77 \pm 0.18$ in IGR\,J00291+5934, indicates radiatively inefficient accretion, albeit with weaker radio emission than tMSPs. Different radio--X-ray normalizations have previously been observed between non-pulsating neutron stars \citep{2016MNRAS.tmp..785T}, and a similar behaviour could occur in pulsating neutron stars.

The correlation index is different than measured over the limited $L_{\rm X} \approx 10^{36}$ to $10^{37}$\,erg\,s\,$^{-1}$ range for the atoll sources in the hard state \citep{2006MNRAS.366...79M, 2016MNRAS.tmp..785T}. This does not, however, preclude IGR\,J00291+5934 and some atoll sources from sharing similar ratiatively inefficient emission mechanisms, especially since IGR\,J00291+5934 has a similar radio--X-ray normalization to the atolls 4U\,1728--34 and Aql\,X--1.

The presence of reflares in SAX\,J1808.4--3658 and lack thereof in IGR\,J00291+5934 \citep[e.g.][]{2011ApJ...726...26H} highlights additional differences between the two sources. However, as SAX\,J1808.4--3658 and IGR\,J00291+5934 are the only two neutron stars with radio coverage in the range $L_{\rm X} \approx 10^{34}$ to $10^{36}$\,erg\,s\,$^{-1}$, no firm conclusion can be drawn as to the causes of their different behaviour.

\section{Conclusions}

For the first time, multi-wavelength observations of SAX\,J1808.4--3658 and IGR\,J00291+5934 have allowed us to study the inflow/outflow coupling within individual neutron stars in the $L_{\rm X} \approx 10^{34}$--$10^{36}$\,erg\,s\,$^{-1}$ range. We find that SAX\,J1808.4--3658 is strongly radio-variable and that it does not follow an obvious $L_{\rm R}$--$L_{\rm X}$ relationship, although at the peak of the outburst it has similar radio and X-ray luminosities as other neutron stars. A combination of factors, such as sparse sampling during the reflares, a long time delay, and strong ejections driven by a propeller, are probably responsible for its behaviour. IGR\,J00291+5934, on the other hand, has a more well-defined radio--X-ray correlation, with $\beta = 0.77 \pm 0.18$. We do not detect IGR\,J17511--3057, but we place stringent limits on its radio emission. The outburst detections and quiescent non-detection of \mbox{Cen\,X--4} are compatible with the radio and X-ray luminosities of other atoll sources. Overall, neutron stars are always at least a factor of five fainter than the radiatively inefficient black hole track \citep{2014MNRAS.445..290G}.

Such varied behaviour across neutron star classes (and even within AMXPs) could be due to the jet launching mechanism, as impacted by the magnetic field and spin of the neutron star. At high X-ray luminosities, where the disk truncates below the corotation radius, there is evidence of different normalizations for each system \citep{2016MNRAS.tmp..785T}. As the mass accretion rate decreases, some systems might enter a strong propeller mode, where matter is ejected more efficiently, increasing the radio flux, possibly contributing to the different correlations seen below $L_{\rm X} \lesssim 10^{36}$\,erg\,s$^{-1}$. It is possible, however, that even without the effects of the propeller mode, neutron stars do not follow a unique radio--X-ray relationship and each system might have a unique behaviour, as suggested by \citet{2014MNRAS.445..290G} for black holes.

The varied behaviour of neutron stars at low X-ray luminosities ($L_{\rm X} \lesssim 0.01 \: L_{\rm Edd}$) is in contrast to black holes, which seem to all settle on a similar radiatively-inefficient track. This emphasises the need to accumulate more data on neutron stars, rather than black holes, since there seem to be several classes of neutron stars. The difficulty of carrying out these observations due to radio faintness, however, could be addressed by the upcoming next-generation radio interferometers, such as MeerKAT \citep{2009IEEEP..97.1522J} and SKA \citep{2009IEEEP..97.1482D}.

We propose that during its next outburst, SAX\,J1808.4--3658 should be monitored approximately twice daily for about three weeks in the X-ray and radio bands. Such dense sampling during the fast X-ray decay and reflares is essential for tracking the disk--jet coupling in the propeller regime, measuring a radio/X-ray time lag, quantifying the differences in the radio spectra at low and high radio luminosities, and ultimately identifying the mechanism behind the radio emission. Even without a radio detection, \mbox{Cen\,X--4} remains a good candidate for future campaigns for testing whether any non-pulsating neutron stars produce jets at the faintest X-ray luminosities ($L_{\rm X} \approx 10^{32}$\,erg\,s$^{-1}$).

\section*{Acknowledgements}

We thank the anonymous referee for constructive feedback that improved the manuscript. VT acknowledges a CSIRS scholarship from Curtin University. JCAM-J is the recipient of an Australian Research Council Future Fellowship (FT140101082). AP and CRD'A are financially supported by an NWO Vidi grant (PI: Patruno). PGJ acknowledges funding from the European Research Council under ERC Consolidator Grant agreement number 647208. TDR acknowledges support from the Netherlands Organisation for Scientific Research (NWO) Veni Fellowship, grant number 639.041.646. RMP acknowledges support from Curtin University through the Peter Curran Memorial Fellowship. RW is supported by a NWO Top Grant, Module 1. This research has made use of NASA's Astrophysics Data System, and of data supplied by the UK Swift Science Data Centre at the University of Leicester. The International Centre for Radio Astronomy Research is a joint venture between Curtin University and the University of Western Australia, funded by the state government of Western Australia and the joint venture partners. The National Radio Astronomy Observatory is a facility of the National Science Foundation operated under cooperative agreement by Associated Universities, Inc. The Faulkes Telescopes are maintained and operated by Las Cumbres Observatory (LCO).




\bibliographystyle{mnras}
\bibliography{radio_ns} 





\bsp	
\label{lastpage}
\end{document}